\documentclass[a4paper,11pt,reqno]{article}
\usepackage{a4wide}
\usepackage{amsmath,amsfonts,amssymb}
\usepackage[english]{babel}
\usepackage{soul}
\usepackage{nicefrac}
\usepackage[mathscr]{euscript}
\usepackage{setspace}
\usepackage{datetime}
\usepackage[nosort]{cite}
\usepackage{mathrsfs}
\usepackage[T1]{fontenc}

\usepackage{txfonts}
\usepackage{kpfonts}
\usepackage{comment}

\usepackage{mathpazo}
\usepackage{graphicx}
\usepackage{multirow}
\usepackage{hhline}

\numberwithin{equation}{section}

\let\oldsqrt\sqrt
\def\sqrt{\mathpalette\DHLhksqrt}
\def\DHLhksqrt#1#2{%
\setbox0=\hbox{$#1\oldsqrt{#2\,}$}\dimen0=\ht0
\advance\dimen0-0.2\ht0
\setbox2=\hbox{\vrule height\ht0 depth -\dimen0}%
{\box0\lower0.4pt\box2}}

\usepackage{hyperref}
\hypersetup{
			colorlinks=true,
			urlcolor=blue,
			citecolor=magenta,
			linkcolor=blue,
     }
     
\usepackage{cite}

\newcommand{\RNum}[1]{\uppercase\expandafter{\romannumeral #1\relax}}

\author{
  \begin{minipage}{.53\linewidth}
    \vspace{1cm}
       \begin{center}
      \begin{small}
               \textbf{Luca Ciambelli} and \textbf{Charles Marteau} 
      \end{small}
    \end{center}
    \vspace{0.5cm}
      \hspace{2.4cm}
{\it \begin{footnotesize}
\hbox{\kern-1.8cm\vbox{\vskip0cm
\begin{center}
 \begin{itemize}
               \item[]CPHT -- Centre de Physique Th\'eorique\\ 
        Ecole Polytechnique, CNRS UMR 7644\\
        Universit\'e Paris--Saclay\\
        91128 Palaiseau Cedex, France
\vskip0.3cm
      \end{itemize}
      \end{center}}}
     \end{footnotesize}}
  \end{minipage}
}

\hypersetup{urlcolor=blue}

\title{\vspace{1.5cm}
 \boldmath 
    \textbf{Carrollian conservation laws and Ricci-flat gravity}
  \unboldmath
}

\begin{document}

\begin{titlepage}
\maketitle
\thispagestyle{empty}

 \vspace{-13,5cm}
  \begin{flushright}
  CPHT-RR101.102018
  \end{flushright}
 \vspace{11.7cm}

\begin{center}
\vspace{1.5cm}
\textsc{Abstract}\\  
\vspace{1.5cm}	
\begin{minipage}{1.0\linewidth}

We construct the Carrollian equivalent of the relativistic energy--momentum tensor, based on variation of the action with respect to the elementary fields of the Carrollian geometry. We prove that, exactly like in the relativistic case, it satisfies conservation equations that are imposed by general Carrollian covariance. In the flat case we recover the usual non-symmetric energy--momentum tensor obtained using N\oe ther procedure. We show how Carrollian conservation equations emerge taking the ultra-relativistic limit of the relativistic ones. We introduce Carrollian Killing vectors and build associated conserved charges. We finally apply our results to asymptotically flat gravity, where we interpret the boundary equations of motion as ultra-relativistic Carrollian conservation laws, and observe that the surface charges obtained through covariant phase-space formalism match the ones we defined earlier.

\end{minipage}
\end{center}


\end{titlepage}

\onehalfspace

\begingroup
\hypersetup{linkcolor=black}
\tableofcontents
\endgroup
\noindent\rule{\textwidth}{0.6pt}

\newpage

\section{Introduction}

The Carroll group was firstly introduced in \cite{Levy1965} as a contraction of the Poincar\'e group for vanishing speed of light and this is referred to as the \emph{ultra-relativistic} limit. The main feature is that, as opposed to the Galilean case, this group allows for boosts only in the time direction: space is absolute.

We could wonder what happens when we take the zero-$c$ limit of a relativistic general-covariant theory. The resulting theory ends up being covariant only under a subset of the diffeormorphisms, as illustrated in \cite{Ciambelli2018} , the so-called \emph{Carrollian diffeormorphisms}
\begin{equation}
t^{\prime}=t^{\prime}(t,\textbf x),\quad
\textbf x^{\prime}=\textbf x^{\prime}(\textbf x).
\label{Cdiff}
\end{equation}
The ultra-relativistic limit breaks the spacetime metric into three independent data, a scalar density, a connection and a spatial metric. These geometric fields are nicely interpreted as constituents of a \emph{Carrollian geometry}, as we will show in Sec. \ref{II}. Now considering an action defined on such a geometry, covariant under \eqref{Cdiff}, we are facing a problem in defining the energy--momentum tensor. Indeed, in general-covariant theories it is obtained as the variation of the action with respect to the metric. This requires the existence of a regular metric \emph{i.e.} of a pseudo-Riemannian manifold, but in the Carrollian case, as we mentioned, there is no spacetime non-degenerate metric. Therefore, we must introduce new objects. The core of Sec. \ref{II} will be dedicated to the definition of these new objects, dubbed \emph{Carrollian momenta}, and obtained as the variation of the action with respect to the $3$ geometric fields mentioned above. 

General covariance usually ensures that the energy--momentum tensor is conserved. In the context of Carroll-covariant theories, we will derive similar conservation equations for the Carrollian momenta. In order to gain confidence with these new definitions, we will study a simple Carrollian action, and show that, on a flat geometrical background, the Carrollian momenta are packaged in a spacetime energy--momentum tensor which coincides with the N\oe ther current associated with spacetime translations. This will be done in Sec. \ref{III}.
 
We will further discuss the intrinsic Carrollian nature of the ultra-relativistic limit. Indeed, in Sec. \ref{IV}, starting from the conservation equations of an energy--momentum tensor, covariant under all changes of coordinates, we reach conservation laws that look strikingly similar to the ones we derived for the Carrollian momenta, which are covariant only under \eqref{Cdiff}. 
 
In general-covariant theories, the existence of a Killing vector allows to build a conserved current by projecting the energy--momentum tensor on the Killing field. This ultimately leads to a conserved charge. After briefly introducing the notion of conserved current in the Carrollian context, we define in Sec. \ref{V} the Carrollian Killing vectors and build their associated currents and charges. 
 
There are by now different instances in which the Carrollian framework has found applications. For instance, it has been used in electromagnetism \cite{Basu2018f} and to discuss the so-called Carroll strings \cite{Cardona2016f}.
The last part of this paper is devoted to yet another application of the Carrollian framework: flat holography. The latter is a holographic correspondence between a theory of asymptotically flat gravity and a non-gravitational theory leaving on its boundary, see \cite{Dappiaggi:2005ci, Barnich:2010eb, Bagchi2013, Fareghbal2014, Hartong2015, Hartong2016, Kapec:2016jld, Ciambelli2018a} for recent progresses in this direction. Asymptotically anti-de-Sitter spacetimes enjoy a timelike pseudo-Riemannian boundary and the associated metric sources its dual operator: the boundary energy--momentum tensor. For asymptotically flat spacetimes, the dual theory leaves on the null infinity $\mathcal{I}^+$. Nevertheless this surface does not carry the same geometrical structure, it is a null hypersurface thus equipped with a Carrollian geometry \cite{Hartong2015} and this will be the source for the Carrollian momenta. The conservation of the latter will be shown to correspond to the gravitational dynamics in the bulk.\footnote{Some attention has been recently given to the interpretation of the bulk dynamics in terms of null conservation laws, see \emph{e.g.} \cite{Hopfmueller2018}.} As a cross check, it has been shown \cite{Duval:2014uva} that the conformal Carroll group has a particular realization which is nothing but the Bondi--Metzner--Sachs (BMS) group \cite{Alessio2018}: the symmetries associated with a Carrollian structure match the asymptotic symmetries of the bulk.

In Secs. \ref{6.1} and \ref{6.2} we focus on the Carrollian theory on $\mathcal{I}^+$ and its relevance for gravitational asymptotically flat duals in 3 and 4 dimensions, and in Sec. \ref{6.3} we study explicit solutions, namely the Robinson--Trautman and the Kerr--Taub--NUT families.
 
\section{Carrollian momenta}\label{II}

We start with a brief reminder on the energy--momentum tensor in the relativistic case, and then define its counterpart, that we call \emph{Carrollian momenta}, on a general Carrollian background. This requires the study of Carrollian geometry and covariance, which will be eventually the guideline for obtaining the conservation equations of these momenta. We also extend our results for a scale invariant theory (Weyl invariant) and write the conservation equations in a Weyl-covariant way. Finally, we focus on the flat case and show how, in this case only, one can promote the Carrollian momenta to a "non-symmetric energy--momentum tensor".

\subsection{A relativistic synopsis}

In a relativistic theory, the energy--momentum tensor is usually defined as
\begin{equation}
T^{\mu\nu}=\frac{-2}{\sqrt{-g}}\frac{\delta S}{\delta g_{\mu\nu}}.
\end{equation}
For a general-covariant theory, it is easy to prove that it is conserved. Indeed, considering the variation of the action under an infinitesimal coordinate transformation $x^{\mu}\rightarrow x^{\mu}+\xi^{\mu}$, we have
\begin{equation}
\delta_{\xi}S=\int \text{d}^{d+1}x\left(\frac{\delta S}{\delta g_{\mu \nu}}\delta_{\xi} g_{\mu \nu}+\frac{\delta S}{\delta \phi}\delta_{\xi} \phi\right)+\text{b. t.},
\end{equation}
where $d+1$ is the spacetime dimension and $\phi$ stands for the various other fields of the theory. We assume that we are on-shell so $\frac{\delta S}{\delta \phi}=0$. Moreover, $\delta_{\xi}$ is the Lie derivative, which for a Levi Civita reads
\begin{equation}
\delta_{\xi}g_{\mu\nu}=\nabla_{\mu}\xi_{\nu}+\nabla_{\nu}\xi_{\mu}.
\end{equation}
We thus obtain
\begin{equation}
\delta_{\xi}S =-\int \text{d}^{d+1}x\sqrt{-g}T^{\mu\nu}\nabla_{\mu}\xi_{\nu}
=\int \text{d}^{d+1}x \sqrt{-g}\nabla_{\mu}T^{\mu\nu}\xi_{\nu}+\text{b. t.}.
\end{equation}
If the theory is general-covariant, $\delta_{\xi}S=0$  for all $\xi$. From this we deduce that $\nabla_{\mu}T^{\mu\nu}$ vanishes on shell, which is the usual conservation law of the energy--momentum tensor.

\subsection{Carrollian geometry}\label{2.2}

We briefly introduce here the Carrollian geometry, as it emerges from an ultra-relativistic ($c\rightarrow 0$) limit of the relativistic metric. It has been shown in \cite{Ciambelli2018, Ciambelli2018a} that the conservation equations of a relativistic energy--momentum tensor, covariant under all diffeomorphisms, lead, in the $c\rightarrow 0$ limit, to equations covariant under a subset called \emph{Carrollian diffeomorphisms} 
\begin{equation}
t^{\prime}=t^{\prime}(t,\textbf x),\quad
\textbf x^{\prime}=\textbf x^{\prime}(\textbf x).
\label{CDiffeo}
\end{equation}
An adequate parametrization for taking this limit is the so-called Randers--Papapetrou, in which the various components transform nicely under this subset of diffeomorphisms. The metric takes the form\footnote{Every metric can be parametrized in this way. The alternative parametrization, known as Zermelo, turns out to be useful for the Galilean limit (see \cite{Ciambelli2018, Duval:2014uoa}).}
\begin{equation}
g=
\begin{pmatrix}
-\Omega^2 & c\Omega b_i \\
c\Omega b_j & a_{ij}-c^2b_ib_j
\end{pmatrix}_{\{c\text{d}t,\text{d}x^i\}} \label{RP}
\end{equation}
where $i=\{1,\dots,d\}$. Indeed, under \eqref{CDiffeo}
\begin{equation}
a^{\prime}_{ij} =a_{kl} J^{-1k}_{\hphantom{-1}i} J^{-1l}_{\hphantom{-1}j} ,\quad
b^{\prime}_{k}=\left( b_i+\frac{\Omega}{J} j_i\right)J^{-1i}_{\hphantom{-1}k},\quad
\Omega^{\prime }=\frac{\Omega}{J},
\label{CTransfo}
\end{equation}
where $J^{k}_{i}=\frac{\partial x^{\prime k}}{\partial x^{i}}$, $j_{i}=\frac{\partial t^{\prime}}{\partial x^{i}}$ and $J=\frac{\partial t^{\prime}}{\partial t}$.  In the $c\rightarrow 0$ limit the metric becomes degenerate, hence we cannot package the different metric fields in a spacetime tensor $g_{\mu\nu}$, but instead we have to treat those three fields separately: time and space decouple as \eqref{CDiffeo} clearly suggests. We therefore trade the metric $g_{\mu\nu}$ for the time lapse $\Omega(t,\textbf{x})$, connection $b_i(t,\textbf{x})$ and spatial metric $a_{ij}(t,\textbf{x})$,\footnote{Hence, we will use $a_{ij}$ to raise and lower spatial indexes in the Carrollian geometry.} which we refer to as Carrollian metric fields, defining a Carrollian geometry. On the derivatives, \eqref{CDiffeo} infers
\begin{equation}
\partial_t^{\prime}=\dfrac{1}{J}\partial_t, \quad \partial_i^{\prime}=J^{-1 k}_i \left(\partial_k- \dfrac{j_k}{J}\partial_t\right),
\end{equation}
which implies that the spatial derivative is not a Carrollian tensor and the temporal one is a density. Therefore we introduce the Carroll-covariant derivatives $\frac{1}{\Omega}\partial_t$ and  $\hat \nabla_i$. In the temporal one the role of $\Omega$ as a time lapse is clear, and the spatial one is defined through its action on scalars as 
\begin{equation}
\hat \partial_i= \partial_i+\dfrac{b_i}{\Omega} \partial_t.
\end{equation}
On Carrollian tensors, it acts as usual with the following Christoffel symbols
\begin{equation}
\hat \gamma^i_{jk}=\dfrac{a^{il}}{2}\left(\hat \partial_j a_{lk}+\hat \partial_k a_{lj}-\hat \partial_l a_{jk}\right).
\end{equation} 
By construction, $\hat \partial_i$ transforms as a Carrollian tensor
\begin{equation}
\hat \partial'_i=J^{-1 k}_i \hat \partial_k,
\end{equation}
and thus we also see clearly the role of $b_i$ as connection.
Out of the Carrollian metric fields, we can build first-order derivative geometrical objects
\begin{eqnarray}
\varphi_i &=& \dfrac{1}{\Omega}\left(\partial_t b_i+\partial_i \Omega\right),\label{acc}\\
\theta &=& \dfrac{1}{\Omega}\partial_t \ln\sqrt{a},\label{exp}\\
\xi_{ij} &=& \dfrac{1}{\Omega}\left(\dfrac{1}{2} \partial_t a_{ij}-\dfrac{1}{d} a_{ij} \partial_t \ln\sqrt{a}\right),\label{shear}\\
\varpi_{ij} &=& \partial_{[i}b_{j]}+\frac{1}{\Omega}b_{[i}\partial_{j]}\Omega+\frac{1}{\Omega}b_{[i}\partial_tb_{j]}.\label{vort}
\end{eqnarray}
They are all Carrollian tensors and they encode the non-flatness of the Carrollian geometrical structure we are defining. They will turn out very useful in writing the conservation equations of the Carrollian momenta defined in the next section.

\subsection{Carrollian momenta}

We define the Carrollian equivalent of the energy--momentum tensor as the three following pieces of data:
\begin{equation}
\mathcal{O}=\frac{1}{\Omega\sqrt{a}}\frac{\delta S}{\delta \Omega},\quad\mathcal{B}^i=\frac{1}{\Omega\sqrt{a}}\frac{\delta S}{\delta b_i}\quad\text{and}\quad\mathcal{A}^{ij}=\frac{1}{\Omega\sqrt{a}}\frac{\delta S}{\delta a_{ij}}. \label{CM}
\end{equation}
Here $\Omega\sqrt{a}$ is the Carrollian counterpart of the relativistic $\sqrt{-g}$ and the variations are taken with respect to the $3$ fields that replace the metric in the Carrollian setting. From now on, we call \eqref{CM} the \emph{Carrollian momenta}.
Before continuing, notice that these quantities transform under Carrollian diffeomorphisms as
\begin{equation}
\mathcal{O}^{\prime}=J \mathcal{O}-\mathcal{B}^i j_i,\quad\mathcal{B}^{i\prime}=J^{i}_j\mathcal{B}^j, \quad\text{and}\quad\mathcal{A}^{ij\prime}=J^i_k J^j_l \mathcal{A}^{kl}.
\end{equation}
The spatial vector $\mathcal{B}^i$ and matrix $\mathcal{A}^{ij}$ are indeed Carrollian tensors. However, $\mathcal{O}$ is not a scalar and, as we will see and use, it is wiser to introduce the scalar combination $\mathcal{E}=\Omega\mathcal{O}+b_i\mathcal{B}^i$.

Given a Carroll-covariant theory, the action is invariant under Carrollian diffeomorphisms, generated by the spacetime vector $\xi$
\begin{equation}
\delta_{\xi}S=0,\quad \xi=\xi^{t}(t,\textbf x)\partial_{t}+\xi^i(\textbf x)\partial_i. \label{diff}
\end{equation}
Notice that $\xi^i$ only depends on $\textbf x$, this is the infinitesimal translation of \eqref{CDiffeo}. Under such an infinitesimal coordinate transformation we have
\begin{equation}
\delta_{\xi}S=\int \text{d}^{d+1}x\left(\frac{\delta S}{\delta \Omega}\delta_{\xi}\Omega+\frac{\delta S}{\delta b_i}\delta_{\xi}b_i+\frac{\delta S}{\delta a_{ij}}\delta_{\xi}a_{ij}+\frac{\delta S}{\delta \phi}\delta_{\xi}\phi\right)+\text{b.t.},
\end{equation}
and the on-shell condition ensures $\frac{\delta S}{\delta \phi}=0$. We need to compute $\delta_{\xi}\Omega$, $\delta_{\xi}b_i$ and $\delta_{\xi}a_{ij}$. In order to do so we compute the infinitesimal version of \eqref{CTransfo}. If $x^{\prime \mu} = x^{\mu}-\xi^{\mu}$, then
\begin{eqnarray}
\delta_{\xi}\Omega &=& \xi\left(\Omega\right)+\Omega\partial_{t}\xi^{t},\label{oit}\\
\delta_{\xi}b_i &=& \xi\left(b_i\right)-\Omega\partial_i\xi^t+b_j\partial_i\xi^j,\label{bit}\\
\delta_{\xi}a_{ij} &=& \xi\left(a_{ij}\right)+\partial_i\xi^ka_{kj}+\partial_j\xi^ka_{ik},
\label{ait}
\end{eqnarray}
where $\xi(f)\equiv \xi^t\partial_tf+\xi^i\partial_if$.
We would like to write these transformations in terms of manifestly Carroll-covariant objects, so we define $X=\Omega\xi^t-b_i\xi^i$. By noticing that the components of a spacetime vector transform as
\begin{equation}
\xi^{t\prime}= J \xi^t+j_i \xi^i, \quad \xi^{i\prime}=J^i_k \xi^k,
\end{equation}
it is straightforward to show that $X$ is the right combination for obtaining a scalar. We thus rewrite \eqref{oit}, \eqref{bit} and \eqref{ait} in terms of $X$, $\xi^i$ and the Carrollian geometrical tensors introduced above
\begin{eqnarray}
\delta_{\xi}\Omega &=& \partial_tX+\Omega\varphi_j\xi^j, \label{reO}\\
\delta_{\xi}b_i &=& -\hat{\partial}_iX+\varphi_iX-2\varpi_{ij}\xi^j+\frac{b_i}{\Omega}\left(\partial_tX+\Omega\varphi_j\xi^j\right), \label{reb}\\
\delta_{\xi}a_{ij} &=& \hat{\nabla}_i\xi_j+\hat{\nabla}_j\xi_i+\frac{X}{\Omega}\partial_ta_{ij}. \label{rea}
\end{eqnarray}
This rewriting hints toward Carrollian covariance, as it replaces $\xi^t$ with $X$. Therefore, we obtain $\delta_{\xi}S=\delta_{X}S+\delta_{\xi^i}S$ with
\begin{eqnarray}
\delta_{X}S &=& \int \text{d}^{d+1}x\Omega\sqrt{a}\left(\mathcal{O}\partial_tX-\mathcal{B}^i\hat{\partial}_iX+\mathcal{B}^i\varphi_iX+\mathcal{B}^i\frac{b_i}{\Omega}\partial_tX+\mathcal{A}^{ij}\frac{X}{\Omega}\partial_ta_{ij}\right),\\
\delta_{\xi^i}S &=& \int \text{d}^{d+1}x\Omega\sqrt{a}\left(\mathcal{O}\Omega\varphi_j\xi^j-2\mathcal{B}^i\varpi_{ij}\xi^j+\mathcal{B}^ib_i\varphi_j\xi^j+2\mathcal{A}^{ij}\hat{\nabla}_i\xi_j\right).
\end{eqnarray}
Finally, demanding $\delta_XS$ and $\delta_{\xi^i}S$ be zero separately and manipulating them, we obtain two conservation equations which are manifestly Carroll-covariant:\footnote{A useful relation is $\mathcal{B}^{i}\hat \partial_i X =-X\left(\hat \nabla_i+\varphi_i\right)\mathcal{B}^{i}$, valid up to total derivatives and for any scalar $X$ and vector ${\cal B}^i$.}
\begin{eqnarray}
\left(\frac{1}{\Omega}\partial_t+\theta\right)\mathcal{E}-\left(\hat{\nabla}_i+2\varphi_i\right)\mathcal{B}^i-\mathcal{A}^{ij}\frac{1}{\Omega}\partial_ta_{ij} &=& 0, \label{cons1}\\
2\left(\hat{\nabla}_i+\varphi_i\right)\mathcal{A}^i_j+2\mathcal{B}^i\varpi_{ij}-\mathcal{E}\varphi_j &=& 0,
\label{Cons2}
\end{eqnarray} 
where we used the already introduced scalar combination $\mathcal{E}=\Omega\mathcal{O}+b_i\mathcal{B}^i$.

Let us briefly summarize. By strict comparison with the relativistic situation, we have defined the momenta of our Carrollian theory to be the variation of the action under the geometrical set of data that characterizes the background. Exploiting the underlying Carrollian symmetry we reached a set of two equations which encode the conservation properties of the momenta. As expected, these equations are fully Carroll-covariant.

\subsection{Weyl covariance}

At the relativistic level, Weyl invariance merges when the theory is invariant under a rescaling $g_{\mu\nu}\rightarrow\frac{g_{\mu\nu}}{\mathcal{B}^2}$ for any $\mathcal{B}$ function of spacetime coordinates.\footnote{This conformal symmetry has important consequences in hydrodynamical holographic theories, \cite{Loganayagam:2008is, Bhattacharyya:2008mz}.} The transformations of $\Omega$, $b_i$ and $a_{ij}$ under Weyl rescaling are deduced from the relativistic Randers--Papapetrou metric \eqref{RP}
\begin{equation}
\Omega\rightarrow\frac{\Omega}{\mathcal{B}},\quad b_i\rightarrow\frac{b_i}{\mathcal{B}}\quad\text{and}\quad a_{ij}\rightarrow\frac{a_{ij}}{\mathcal{B}^2}.
\label{WeylTransf}
\end{equation}
If the action is invariant under such transformations,
\begin{equation}
\delta_{\lambda} S=\int \text{d}^{d+1}x\Omega\sqrt{a}\left(\mathcal{O}\delta_{\lambda}\Omega+\mathcal{B}^i\delta_{\lambda} b_i+\mathcal{A}^{ij}\delta_{\lambda} a_{ij}\right)=\int \text{d}^{d+1}x \Omega\sqrt{a}\lambda\left(\mathcal{O}\Omega+\mathcal{B}^i b_i+2\mathcal{A}^{ij} a_{ij}\right)
\end{equation}
has to vanish for every $\lambda(t,\textbf x)$. Therefore
\begin{equation}
\delta_{\lambda} S=0 \quad \Rightarrow \quad \mathcal{E}=-2\mathcal{A}^i_i. \label{traceless}
\end{equation}
We will refer to this condition as the \emph{conformal state equation}, it is the equivalent of the tracelessness of the energy--momentum tensor in the relativistic case. From \eqref{WeylTransf} we deduce the following transformations of the Carrollian momenta
\begin{equation}
\mathcal{O}\rightarrow\mathcal{B}^{d+2}\mathcal{O}, \quad\mathcal{B}^i\rightarrow\mathcal{B}^{d+2}\mathcal{B}^i\quad\text{and}\quad\mathcal{A}^{ij}\rightarrow\mathcal{B}^{d+3}\mathcal{A}^{ij}.
\end{equation}
This implies also $\mathcal{E}\rightarrow\mathcal{B}^{d+1}\mathcal{E}$. 

We would like to write the conservation equations in a manifestly Weyl-covariant form. To do so, we decompose $\mathcal{A}^{ij}=-\frac{1}{2}\left(\mathcal{P}a^{ij}-\Xi^{ij}\right)$ with $\Xi^{ij}$ traceless, such that the constraint \eqref{traceless} becomes $\mathcal{E}=d\mathcal{P}$.  This enable us rewriting \eqref{cons1} and \eqref{Cons2} as
\begin{eqnarray}
\left(\frac{1}{\Omega}\partial_t+\frac{d+1}{d}\theta\right)\mathcal{E}-\left(\hat{\nabla}_i+2\varphi_i\right)\mathcal{B}^i-\Xi^{ij}\xi_{ij} &=& 0, \label{WCE}\\
\left(\hat{\nabla}_i+\varphi_i\right)\Xi^i_j-\dfrac{1}{d}\left(\hat{\partial}_j+(d+1)\varphi_j\right)\mathcal{E}+2\mathcal{B}^i\varpi_{ij} &=& 0. \label{WCEbis}
\end{eqnarray}
The Carrollian derivatives are not covariant under Weyl rescaling, since the latter brings extra shift terms. In order to reach manifestly Weyl-Carroll-covariant equations, we can upgrade the Carroll derivatives to Weyl-Carroll ones. Among the Carrollian first derivative tensors introduced above, $\varphi_i$ and $\theta$ are Weyl connections as
\begin{equation}
 \label{weyl-geometry-2-abs}
\varphi_{i}\to \varphi_{i}-\hat\partial_i\ln \mathcal{B},\quad \theta\to \mathcal{B}\theta-\frac{d}{\Omega}\partial_t \mathcal{B}.
\end{equation} 
Therefore, they can be used for defining the Weyl-Carroll derivative. For a weight-$w$ scalar function  $\Phi$, \emph{i.e.} a function scaling with $\mathcal{B}^w$ under Weyl, and a weight-$w$ vector, the Weyl-Carroll spatial and temporal derivatives are defined as
\begin{eqnarray}
\label{CWs-Phi}
\hat{\mathscr{D}}_j \Phi &=& \hat\partial_j \Phi +w \varphi_j \Phi, \\
\frac{1}{\Omega}\hat{\mathscr{D}}_t \Phi &=&
\frac{1}{\Omega}\partial_t \Phi +\frac{w}{d} \theta \Phi, \\
\hat{\mathscr{D}}_j V^l &=& \hat\nabla_j V^l +(w-1) \varphi_j V^l +\varphi^l V_j -\delta^l_j V^i\varphi_i, \\
\frac{1}{\Omega}\hat{\mathscr{D}}_t V^l &=&
\frac{1}{\Omega}\partial_t V^l +\frac{w}{d} \theta V^l
+\xi^{l}_{\hphantom{l}i} V^i,
\end{eqnarray}
such that under a Weyl transformation
\begin{eqnarray}
\hat{\mathscr{D}}_j \Phi &\to& \mathcal{B}^w \hat{\mathscr{D}}_j \Phi, \\
\frac{1}{\Omega}\hat{\mathscr{D}}_t \Phi &\to& \mathcal{B}^{w+1}\frac{1}{\Omega}\hat{\mathscr{D}}_t \Phi, \\
\hat{\mathscr{D}}_j V^l &\to& \mathcal{B}^w\hat{\mathscr{D}}_j V^l, \\
\frac{1}{\Omega}\hat{\mathscr{D}}_t V^l &\to& \mathcal{B}^{w+1}\frac{1}{\Omega}\hat{\mathscr{D}}_t V^l.
\end{eqnarray}
The action on any other tensor is obtained using the Leibniz rule. 

Eventually, we can write \eqref{WCE} and \eqref{WCEbis} using these derivatives as
\begin{eqnarray}
\frac{1}{\Omega}\hat{\mathscr{D}}_t\mathcal{E}-\hat{\mathscr{D}}_i\mathcal{B}^i-\Xi^{ij}\xi_{ij} &=& 0,\\
-\dfrac{1}{d}\hat{\mathscr{D}}_j\mathcal{E}+2\mathcal{B}^i\varpi_{ij}+\hat{\mathscr{D}}_i\Xi^i_j &=& 0.
\end{eqnarray}
Not only these equations are now very compact, they are also manifestly Weyl-Carroll-covariant.

\subsection{The flat case}\label{2.5}

So far we have worked on general Carrollian geometry, \emph{i.e.} we did not impose any particular value of $\Omega$, $b_i$ and $a_{ij}$. We now restrict our attention to the flat Carrollian background.\footnote{We refer here to flat Carrollian geometry as the geometry for which the Carroll group is an isometry, see \emph{e.g.} \cite{Duval:2014uoa}.} 

At the relativistic level, the Poincar\'e group is defined as the set of coordinate transformations that leave the Minkowski metric invariant. By strict analogy, the Carroll group is defined as the set of transformations that preserve the Carrollian flatness, \cite{Duval:2014uoa}. Therefore, the Carroll group corresponds to the transformations satisfying
\begin{equation}
\partial_t\rightarrow\partial_t,\quad\delta_{ij}\text{d}x^i\text{d}x^j\rightarrow\delta_{ij}\text{d}x^i\text{d}x^j,\quad b_{0i}\rightarrow R^j_i\left(b_{0j}+\beta_j\right),
\end{equation}
with $b_{0i}$ constant. The resulting change of coordinates is
\begin{equation}
t^{\prime}=t+\beta_ix^i+t_0,\quad x^{\prime i}=R^i_jx^j+x_0^i,
\label{FlatC}
\end{equation}
where $t_0\in\mathbb{R}$, $\{x_0^i,\beta_i\}\in\mathbb{R}^{d}$ and $R^i_j\in\text{O}(d)$. This group is known in the literature as the Carroll group. \footnote{The Carroll group was already shown to be the symmetry group of flat zero signature geometries in the precursory work \cite{Marc}.}

Recasting \eqref{cons1} and \eqref{Cons2} for $a_{ij}(t,\textbf x)=\delta_{ij}$, $\Omega(t,\textbf x)=1$ and $b_i(t,\textbf x)=b_{0i}$, we obtain
\begin{eqnarray}
\partial_t\mathcal{O}-\partial_i\mathcal{B}^i &=& 0, \label{flat1}\\
2\partial_i\mathcal{A}^i_j+2b_{0i}\partial_t\mathcal{A}^i_j &=& 0 \label{flat2}.
\end{eqnarray}
The momenta appearing in these two equations can be packaged in a spacetime energy--momentum tensor (where spacetime does not mean relativistic)
\begin{equation}
T^{\mu\nu}=
\begin{pmatrix}
\mathcal{O} & -2b_{0k}\mathcal{A}^{ki}\\
-\mathcal{B}^j&-2\mathcal{A}^{ij}
\end{pmatrix}.
\label{EMflat}
\end{equation}
The usual conservation of this tensor $\partial_{\mu}T^{\mu\nu}=0$ is ensured by the conservation equations of the momenta, namely \eqref{flat1} and \eqref{flat2}. This tensor is not symmetric, but this should not come as a surprise: it is not defined throughout the variation of the action with respect to the spacetime metric (symmetric by construction), instead it is defined using the Carrollian metric fields.\footnote{Although the construction is different, another example of non-symmetric Carrollian energy--momentum tensor can be found in \cite{Boer2018}.} Finally notice that this spacetime lifting procedure was possible here due to the flatness of the Carrollian geometry. In general backgrounds, this is not possible, and the very concept of spacetime energy--momentum tensor is ambiguous--whereas the Carrollian momenta are by construction well suited. 

As a conclusive remark notice that the Carroll group contains spacetime translations, so if a theory is invariant under this group, there will be a set of $d+1$ N\oe ther currents associated with spacetime translations. Packaging them in a $d+1$-dimensional kind of N\oe ther energy--momentum tensor, enables us comparing it with \eqref{EMflat}, as we do in the next section.

\section{A Carrollian scalar-field action}\label{III}

In order to probe our results, we start with the example of a single scalar field $\phi(t,\textbf x)$. We begin the study on a general Carrollian background and show that the momenta are conserved. Then, we restrict the geometry to the flat case, where spacetime translational invariance of the theory allows us to compare our energy--momentum tensor (defined only in the flat case, as in Sec. \ref{2.5}) to the conserved current computed using N\oe ther procedure. The two energy--momentum tensors will turn out to be equivalent up to divergence-free terms. 

In order to ensure Carroll invariance of the scalar-field action, we need to trade the usual derivatives for the Carrollian ones. So we consider the action
\begin{equation}
S[\phi]=\frac{1}{2}\int \text{d}^{d+1}x\Omega\sqrt{a}a^{ij}\hat{\partial}_i\phi\hat{\partial}_j\phi=\int \text{d}^{d+1}x\mathcal{L}, \label{act}
\end{equation}
which is manifestly covariant. The equations of motion are readily determined 
\begin{equation}
\left(\hat{\nabla}_i+\varphi_i\right)\hat{\partial}^i\phi=0. \label{eom}
\end{equation}
The Carrollian momenta are 
\begin{eqnarray}
\mathcal{E} &=& \frac{1}{2}\hat{\partial}_i\phi\hat{\partial}^i\phi, \label{m1}\\
\mathcal{B}^i &=& \frac{1}{\Omega}\partial_t\phi\hat{\partial}^i\phi, \label{m2}\\
\mathcal{A}^{ij} &=& \frac{1}{2}\left(\frac{1}{2}a^{ij}\hat{\partial}^k\phi\hat{\partial}_k\phi-\hat{\partial}^i\phi\hat{\partial}^j\phi\right). \label{m3}
\end{eqnarray}
These momenta are conserved on shell since the conservation equations \eqref{cons1} and \eqref{Cons2} are automatically satisfied given the equations of motion \eqref{eom}. This last result shows unambiguously the relevance of these objects. Notice moreover that these momenta satisfy the conformal state equation \eqref{traceless} only for $d=1$. In fact this action can be recovered from an ultra-relativistic limit of the free relativistic scalar theory, which is known to be conformal only in $2$ spacetime dimensions.

We now impose the Carrollian background to be flat. In this case, the action \eqref{act} becomes
\begin{equation}
S[\phi]=\int \text{d}^{d+1}x\mathcal{L}=\frac{1}{2}\int \text{d}^{d+1}x\delta^{ij}\left(\partial_i+b_{0i}\partial_t\right)\phi\left(\partial_j+b_{0j}\partial_t\right)\phi,\label{flatf}
\end{equation}
which is invariant under spacetime translations. In the flat case, we can lift the Carrollian momenta into a spacetime energy--momentum tensor \eqref{EMflat}, which here takes the form
\begin{equation}
T^{\mu\nu}=
\begin{pmatrix}
\frac{1}{2}\hat{\partial}_i\phi\hat{\partial}^i\phi-b_{0i}\partial_t\phi\hat{\partial}^i\phi & -\frac{b_0^i }{2}\hat{\partial}^k\phi\hat{\partial}_k\phi+b_{0k}\hat{\partial}^k\phi\hat{\partial}^i\phi\\
-\partial_t\phi\hat{\partial}^i\phi & -\frac{1}{2}a^{ij}\hat{\partial}^k\phi\hat{\partial}_k\phi+\hat{\partial}^i\phi\hat{\partial}^j\phi
\end{pmatrix},
\label{flatT}
\end{equation}
and it is conserved. 

The action \eqref{flatf} is invariant under spacetime translations. As stated in the previous section, we therefore have $d+1$ associated N\oe ther currents 
\begin{equation}
\hat{T}^{\mu\nu}=\frac{\partial \mathcal{L}}{\partial\partial_{\mu}\phi}\partial^{\nu}\phi-\eta^{\mu\nu}\mathcal{L},
\end{equation}
which explicitly read:
\begin{eqnarray}
\hat{T}^{tt} &=& \frac{1}{2}\hat{\partial}_i\phi\hat{\partial}^i\phi-b_{0i}\hat{\partial}^i\phi\partial_t\phi,\label{NoetherEMtt}\\
\hat{T}^{it} &=& -\hat{\partial}^i\phi\partial_t\phi,\label{NoetherEMit}\\
\hat{T}^{ti} &=& b_{0j}\hat{\partial}^j\phi\partial^i\phi,\label{NoetherEMti}\\
\hat{T}^{ij} &=& \hat{\partial}^i\phi\partial^j\phi-\frac{1}{2}\delta^{ij}\hat{\partial}^k\phi\hat{\partial}_k\phi.
\label{NoetherEMij}
\end{eqnarray}
The conservation $\partial_{\mu}\hat{T}^{\mu\nu}=0$, is achieved thanks to the equations of motion \eqref{eom} for flat geometry $\hat{\partial}^i\hat{\partial}_i\phi=0$. 

We can now compare the energy--momentum tensor \eqref{flatT} with \eqref{NoetherEMtt}, \eqref{NoetherEMit}, \eqref{NoetherEMti} and \eqref{NoetherEMij}. We obtain
\begin{equation}
\hat{T}^{\mu\nu}=T^{\mu\nu}+B^{\mu\nu},
\end{equation}
with
\begin{eqnarray}
B^{tt} &=& 0, \\
B^{it} &=& 0,\\
B^{ti} &=& -b_{0}^ib_0^j\hat{\partial}_j\phi\partial_t\phi+\frac{1}{2}b_0^i\hat{\partial}_k\phi\hat{\partial}^k\phi,\\
B^{ij} &=& -b_0^j\partial_t\phi\hat{\partial}^i\phi.
\end{eqnarray} 
As anticipated, the tensor $B^{\mu\nu}$ is divergenceless on-shell $\partial_{\mu}B^{\mu\nu}=0$, which implies that the two energy--momentum tensors carry the same physical information on the theory. 

\section{Ultra-relativistic limit: the emergence of Carrollian physics}\label{IV}

In the previous sections, we have intrinsically defined the Carrollian momenta starting from the metric fields of a Carrollian geometry. The Carrollian geometry was inspired by an ultra-relativistic contraction of the relativistic metric. We will see now how the ultra-relativistic limit can be directly taken at the level of the conservation equation of the relativistic energy--momentum tensor. This limit provides a richer structure, with more equations and fields. This is neither surprising nor contradictory. It is suggested by the dual Galilean limit, \cite{Ciambelli2018a}. Indeed, in the non-relativistic case, on top of the momentum and energy conservation, an extra equation arises, which is ultimately identified with the continuity equation. A similar phenomenon occurs in the Carrollian case: additional fields and equations survive in the limit, and this is controlled by our choice of $c$-dependence of the fields. 

Given a vector field $u^{\mu}$, normalized as $u^2=-c^2$ with respect to the relativistic metric \eqref{RP}, the energy--momentum tensor can always be decomposed as\footnote{Reminder of the conventions: $x^{\mu}=(x^0,x^i)=(ct,x^i)$.}
\begin{equation}
T^{\mu\nu}=\left(\mathcal{E}+\mathcal{P}\right)\frac{u^{\mu}u^{\nu}}{c^2}+\mathcal{P}g^{\mu\nu}+\tau^{\mu\nu}+\frac{q^{\mu}u^{\nu}}{c^2}+\frac{q^{\nu}u^{\mu}}{c^2}. \label{EM}
\end{equation}
In the hydrodynamic interpretation, $\mathcal{E}$ and $\mathcal{P}$ are the energy density and pressure of the fluid, $g^{\mu\nu}$ is the spacetime metric and $\tau^{\mu\nu}$ and $q^{\mu}$ are the transverse dissipative tensors, named viscous stress tensor and heat current. We choose to adapt the velocity to the geometry $u^{\mu}=\left(\frac{c}{\Omega},0\right)$: the fluid is at rest. The advantage of this choice is that the dissipative tensors, since transverse, have only spatial independent components. Inspired by flat holography \cite{Ciambelli2018a}, we choose a particular scaling of these tensors in $c$, namely
\begin{equation}
\tau^{ij}=-\frac{\Sigma^{ij}}{c^2}-\Xi^{ij}\quad\text{and}\quad q^i=-\mathcal{B}^i+c^2\pi^i.
\end{equation}
A more general dependence could have been considered. This would add new fields and new equations to the resulting Carrollian theory, whereas the present choice will be sufficient for the examples we want to analyze. Notice that the $c$-independent situation is recovered for $\Sigma^{ij}=0=\pi^i$. 
We now perform the zero-$c$ limit of $\nabla_{\mu}T^{\mu\nu}=0$. Defining again $\mathcal{A}^{ij}=-\frac{1}{2}\left(\mathcal{P}a^{ij}-\Xi^{ij}\right)$, we obtain the following set of equations\footnote{This limit is performed using the decomposition \eqref{EM} and the Randers--Papapetrou parametrization of the spacetime metric. For the detailed derivation of these equations, see \cite{Ciambelli2018}.}
\begin{eqnarray}
\left(\frac{1}{\Omega}\partial_t+\theta\right)\mathcal{E}-\left(\hat{\nabla}_i+2\varphi_i\right)\mathcal{B}^i-\mathcal{A}^{ij}\frac{1}{\Omega}\partial_ta_ {ij} = 0,& \label{RC1}\\
2\left(\hat{\nabla}_i+\varphi_i\right)\mathcal{A}^i_j+2\mathcal{B}^i\varpi_{ij}-\mathcal{E}\varphi_j-\left(\frac{1}{\Omega}\partial_t+\theta\right)\pi_j = 0,& \label{RC2}\\
\left(\frac{1}{\Omega}\partial_t+\theta\right)\mathcal{B}_j+\left(\hat{\nabla}_i+\varphi_i\right)\Sigma^i_j = 0,& \label{RC3}\\
\Sigma^{ij}\xi_{ij}+\frac{\theta}{d}\Sigma^i_i = 0. &
\label{RC4}
\end{eqnarray}
As advertised, we immediately recognize \eqref{RC3} and \eqref{RC4} as the Carrollian counterpart of the continuity equation: these are two consistency equations of the limit. Notice moreover how these equations reduce to the Carrollian equations \eqref{cons1} and \eqref{Cons2} when the dissipative terms have no $c$-dependence, $\Sigma^{ij}=0=\pi^i$, together with the additional constraint $\left(\frac{1}{\Omega}\partial_t+\theta\right)\mathcal{B}_j=0$. This result undoubtedly shows the nature of the ultra-relativistic limit: it is a Carrollian limit. Conversely, this analysis gives credit to our intrinsic Carrollian construction of the previous sections. 

Summarizing, we have shown how the ultra-relativistic expansion gives rise to a leading Carrollian behavior. Furthermore, we have analyzed the extra inputs this limit brings and the associated conservation equations. It is remarkable how the Carrollian momenta intrinsically defined using Carrollian geometry match the ultra-relativistic limit. 

We conclude with an aside important remark: we have taken the ultra-relativistic limit of the conservation equations because it would have been inconsistent to compute directly the limit of the energy--momentum tensor itself. Indeed we would have lost information on the fields which survive and the conservation equations they satisfy. This confirms that we have to give up the concept of spacetime energy--momentum tensor on general Carrollian backgrounds, as anticipated in \cite{Ciambelli2018} but sometimes disregarded in the current literature.

\section{Charges}\label{V}

This section is dedicated to the definition of charges in the Carrollian framework. Charges are conserved quantities associated with a symmetry of the theory. Relativistically, the latter can be generated by a Killing vector field. By projecting the energy--momentum tensor on the Killing vector, we obtain a conserved current. We will show here how to implement this procedure in the Carrollian case. In order to do so, we firstly derive charges starting from a conserved Carrollian current. Secondly, we define Carrollian Killing and conformal Killing vectors. Thirdly, we build conserved charges associated with conformal Killing vectors. This will be useful for the forthcoming examples involving asymptotically flat gravity. Finally, we give another example of Carrollian action and compute the charges to illustrate our results.

\subsection{Conserved Carrollian current and associated charge}\label{gench}

We show here a way to define a conserved charge starting from a conserved current. In this derivation we never impose the current to be associated with a Killing vector, therefore our construction is very general.
Whenever we have a scalar $\mathcal{J}$ and a vector $\mathcal{J}^i$ satisfying
\begin{equation}
\left(\frac{1}{\Omega}\partial_t+\theta\right)\mathcal{J}+\left(\hat{\nabla}_i+\varphi_ i\right)\mathcal{J}^i=0,\label{consc}
\end{equation}
we can build the conserved charge
\begin{equation}
\mathcal{Q}=\int_{\Sigma_t}\text{d}^dx\sqrt{a}\left(\mathcal{J}+b_ i\mathcal{J}^i\right),
\label{Charge}
\end{equation}
where $\Sigma_ t$ is a constant-time slice. A way to derive this formula is to start from the relativistic level: consider a conserved current $J^\mu$, the charge is then
\begin{equation}
Q=\int_{\Sigma_ t}\text{d}^dx\sqrt{\sigma}n_{\mu}J^{\mu}.
\end{equation}
Here $n_{\mu}$ is the unit vector normal to $\Sigma_t$ and $\sigma_{\mu\nu}$ is the induced metric on $\Sigma_t$. In order to perform the zero-$c$ limit, we decompose $J^{\mu}$ in an already Carroll-covariant basis
\begin{equation}
J=\mathcal{J}\left(\frac{c}{\Omega}\partial_0\right)+\mathcal{J}^{i}\left(\partial_i+\frac{cb_i}{\Omega}\partial_0\right).
\end{equation}
Then, using the Randers--Papapetrou parametrization for the relativistic spacetime metric   $\text{d}s^2=-c^2(\Omega \text{d}t - b_i \text{d}x^i)^2+a_{ij}\text{d}x^i\text{d}x^j$, we obtain
\begin{equation}
\sqrt{\sigma}=\sqrt{a}+\mathcal{O}\left(c^2\right),\quad
n_0=c\Omega+\mathcal{O}\left(c^3\right),\quad
J^0=\frac{c}{\Omega}\left(\mathcal{J}+b_i\mathcal{J}^i\right).
\end{equation} 
Therefore, we find $Q\underset{c\rightarrow0}{\rightarrow}c^2\mathcal{Q}$, showing the relevance of the proposed Carrollian charge \eqref{Charge}.

\subsection{Carrollian Killing vectors and associated conserved currents}\label{5.2}

A Killing vector is usually defined as a vector field that preserves the metric. Analogously, we define the Carrollian Killing vector $\xi$ to be the vector satisfying\footnote{This is the translation in our language of ${\cal L}_{X}g=0$ and ${\cal L}_X\xi=0$ of (III.6) in \cite{Duval:2014uoa}.}
\begin{equation}
\delta_{\xi}\Omega=0=\delta_{\xi}a_{ij},
\end{equation} 
where $\delta_{\xi}$ is the Lie derivative. This gives rise to two Killing equations on $\xi$, which are exactly \eqref{reO} and \eqref{rea},\footnote{On top of these equations, a Carrollian Killing vector has a time independent spatial part, \emph{i.e.} $\partial_t \xi^i=0$.}
\begin{eqnarray}
\partial_tX+\Omega\varphi_j\xi^j &=& 0,\label{CK1}\\
\hat{\nabla}_i\xi_j+\hat{\nabla}_j\xi_i+\frac{X}{\Omega}\partial_ta_{ij} &=& 0,
\label{CK2}
\end{eqnarray}
where we recall $X=\Omega\xi^t-b_i\xi^i$. Notice that these equations do not actually depend on $b_i$.

The generalization to conformal Carrollian Killing vectors is straightforward. We call $\xi$ a conformal Carrollian Killing vector if
\begin{equation}
\delta_{\xi}\Omega=\lambda\Omega\quad\text{and}\quad\delta_{\xi}a_{ij}=2\lambda a_{ij}.
\end{equation} 
It obeys the following conformal Killing equations:
\begin{eqnarray}
\partial_tX+\Omega\varphi_j\xi^j &=& \lambda\Omega,\label{CCK1}\\
\hat{\nabla}_i\xi_j+\hat{\nabla}_j\xi_i+\frac{X}{\Omega}\partial_ta_{ij} &=& 2\lambda a_{ij}. \label{CCK2}
\end{eqnarray}
In particular from the last equation we obtain $\lambda=\frac{1}{d}\left(\hat{\nabla}_i\xi^i+\frac{X}{\Omega}\partial_t\ln\sqrt{a}\right)$. This general construction is very useful, as we will shortly confirm.

We now build a conserved current by projecting the Carrollian momenta on a Carrollian Killing vector, exactly like in the relativistic case. Indeed consider the following Carrollian current:
\begin{equation}
\mathcal{J}=\xi_i\mathcal{B}^i, \quad\mathcal{J}^i=\xi_j\Sigma^{ij}.
\end{equation}
It is conserved provided $\xi$ satisfies \eqref{CK2}, and the Carrollian conservation equations \eqref{RC3} and \eqref{RC4} are verified. According to Sec. \ref{gench}, the corresponding conserved charge is
\begin{equation}
\mathcal{Q}_{\xi}=\int_{\Sigma_t}\text{d}^dx\sqrt{a}\xi_i\left(\mathcal{B}^i+b_ j\Sigma^{ji}\right),
\label{charge2}
\end{equation}
This charge is also conserved when $\xi$ satisfies \eqref{CCK2}, if we further impose the condition $\Sigma^i_i=0$.

\boldmath\subsection{Charges for $\mathcal{B}^i=0$}\unboldmath\label{5.3}

We will show in Sec. \ref{VI} that the equations describing the dynamics of asymptotically flat spacetimes in 3 and 4 dimensions can be related to Carrollian conservation laws for $\mathcal{B}^i=0$. For this reason we focus here on this particular case and build other conserved currents associated with conformal Killing vectors. In Sec. \ref{VI} we will observe that the corresponding charges match the surface charges obtained through covariant phase-space formalism. 

The Carrollian conservation equations obtained from the ultra-relativistic limit \eqref{RC1} and \eqref{RC2}, for $\mathcal{B}^i=0$, become
\begin{eqnarray}
\left(\frac{1}{\Omega}\partial_t+\theta\right)\mathcal{E}-\mathcal{A}^{ij}\frac{1}{\Omega}\partial_ta_ {ij} &=& 0, \label{RC11}\\
2\left(\hat{\nabla}_i+\varphi_i\right)\mathcal{A}^i_j-\mathcal{E}\varphi_j-\left(\frac{1}{\Omega}\partial_t+\theta\right)\pi_j &=& 0. \label{RC22}
\end{eqnarray}
We could have also reported the two equations on $\Sigma^{ij}$, \eqref{RC3} and \eqref{RC4}, but they are immaterial here. Consider a Killing vector $\xi$, the following charge, up to boundary terms, is conserved
\begin{equation}
{\cal C}_{\xi}=\int_{\Sigma_t}\text{d}^dx\sqrt{a}\left(X\mathcal{E}-\xi^i\pi_i+2b_i\xi^j\mathcal{A}^i_j\right),
\label{Charge1}
\end{equation}
assuming only \eqref{RC11} and \eqref{RC22}. This charge is also conserved when $\xi$ is a conformal Killing vector, if we further impose the conformal state equation $\mathcal{E}=-2\mathcal{A}^i_i$. According to Sec. \ref{gench}, the corresponding conserved current reads\footnote{Its conservation \eqref{consc} is ensured thanks to the Killing equations together with \eqref{RC11} and \eqref{RC22}.}
\begin{equation}
\mathcal{J}=X\mathcal{E}-\xi^i\pi_i, \quad\mathcal{J}^i=2\xi^j\mathcal{A}^i_j.
\end{equation}

It is interesting to investigate the flat case $a_{ij}(t,\textbf x)=\delta_{ij}$, $\Omega(t,\textbf x)=1$ and $b_i(t,\textbf x)=b_{0i}$. Here, \eqref{RC11} and \eqref{RC22} can be written as $\partial_{\mu}T^{\mu\nu}=0$ with\footnote{We recall that for $\mathcal{B}^i=0$, $\mathcal{E}=\Omega\mathcal{O}$. Thus in the flat case $\mathcal{E}=\mathcal{O}$.}
\begin{equation}
T^{\mu\nu}=\begin{pmatrix}
\mathcal{O} & -2b_{0k}\mathcal{A}^{ki}+\pi^i\\
0&-2\mathcal{A}^{ij}
\end{pmatrix},
\label{flatstress}
\end{equation}
and we notice that the charge, up to a divergenceless term, takes the usual form
\begin{equation}
\mathcal{C}^{\text{Flat}}_{\xi}=\int_{\Sigma_t}\text{d}^dx\left(\xi^t\mathcal{O}-\xi^ib_{0i}\mathcal{O}-\xi^i\pi_i+2b_{0i}\xi^j\mathcal{A}^i_j\right)=-\int_{\Sigma_t}\text{d}^dxT^{0\mu}\xi_{\mu}+\tilde{\mathcal{C}}_{\xi^i},
\end{equation}
with $\tilde{\mathcal{C}}_{\xi^i}=-\int_{\Sigma_t}\text{d}^dx\xi^ib_{0i}\mathcal{O}$, which is separately conserved. 

For $\xi$ and $\eta$ Killing vectors, we define the brackets
\begin{equation}
\begin{split}
&\{\mathcal{Q}_{\xi},\mathcal{Q}_{\eta}\}\equiv\int_{\Sigma_t} \text{d}^dx\delta_{\eta}\left[\sqrt{a}\xi_i\left(\mathcal{B}^i+b_ j\Sigma^{ji}\right)\right],\\
&\{\mathcal{C}_{\xi},\mathcal{C}_{\eta}\}\equiv\int_{\Sigma_t} \text{d}^dx\delta_{\eta}\left[\sqrt{a}\left(X\mathcal{E}-\xi^i\pi_i+2b_i\xi^j\mathcal{A}^i_j\right)\right].
\end{split}
\end{equation}
Here $\delta_{\eta}$ is the Lie derivative acting on the metric fields and the momenta, but not on $\xi^t$ and $\xi^i$. A lengthly computation (see appendix \ref{App}) shows that the charges $\mathcal{Q}_{\xi}$ and $\mathcal{C}_{\xi}$ equipped with these brackets form two representations of the Carrollian Killing algebra:
\begin{equation}
\begin{split}
\{\mathcal{Q}_{\xi},\mathcal{Q}_{\eta}\}=\mathcal{Q}_{[\xi,\eta]}\quad\text{and}\quad
\{\mathcal{C}_{\xi},\mathcal{C}_{\eta}\}=\mathcal{C}_{[\xi,\eta]}.
\end{split}
\end{equation}
We can extend these results to the conformal Killing algebra when imposing the conformal state equation $\mathcal{E}=-2\mathcal{A}^i_i$ for the charge $\mathcal{C}_{\xi}$ and the condition $\Sigma^i_i=0$ for the charge $\mathcal{Q}_{\xi}$.

\subsection{Application to the scalar field}

We close this section with an example of scalar-field action whose Carrollian momenta reproduce exactly the conservation equations described in Sec. \ref{5.3}. Consider a scalar field $\phi(t,\textbf x)$ and the following Carroll-covariant action:
\begin{equation}
S\left[\phi\right]=\frac{1}{2}\int \text{d}^{d+1}x \sqrt{a}\dfrac{\dot{\phi}^2}{\Omega}=\int \text{d}^{d+1}x\mathcal{L},
\end{equation}
where $\dot{\phi}=\partial_t\phi$. The equation of motion reads
\begin{equation}
\left(\frac{1}{\Omega}\partial_t+\theta\right)\left(\frac{\dot{\phi}}{\Omega}\right)=0,
\end{equation}
and we find the following Carrollian momenta through the variational definition \eqref{CM}
\begin{equation}
\mathcal{E}=-\frac{1}{2\Omega^2}\dot{\phi}^2,\quad\mathcal{B}^i=0\quad\text{and}\quad\mathcal{A}^{ij}=\frac{1}{4\Omega^2}\dot{\phi}^2a^{ij}.
\end{equation}
Carrollian conservation equations of the type \eqref{RC11} and \eqref{RC22} are satisfied provided $\pi_i=\frac{1}{\Omega}\dot{\phi}\hat{\partial}_i\phi$. In the flat case the energy--momentum tensor \eqref{flatstress} computed earlier becomes:
\begin{equation}
T^{\mu\nu}=
\begin{pmatrix}
-\frac{1}{2}\dot{\phi}^2 & \frac{1}{2}b_{0}^i\dot{\phi}^2+\dot{\phi}\partial^i\phi\\
0&-\frac{1}{2}\dot{\phi}^2\delta^{ij}
\end{pmatrix}.
\end{equation}
As in the other example of scalar-field action (Sec. \ref{III}), this object coincides with the N\oe ther current associated with spacetime translations, up to a divergenceless term.

We can now focus on the charges in the Hamiltonian formalism. Defining the conjugate momentum $\psi=\dfrac{\partial \mathcal{L}}{\partial \dot{\phi}}=\dfrac{\sqrt{a}}{\Omega}\dot{\phi}$, and writing the Carrollian momenta in terms of $\phi$ and $\psi$, we obtain
\begin{equation}
\mathcal{E}=-\frac{1}{2}\left(\frac{\psi}{\sqrt{a}}\right)^2, \quad \pi_i=\frac{\psi}{\sqrt{a}}\left(\partial_i\phi+b_i\frac{\psi}{\sqrt{a}}\right) \quad \text{and} \quad \mathcal{A}^{ij}=\frac{1}{4}\left(\frac{\psi}{\sqrt{a}}\right)^2a^{ij}.
\end{equation}
Therefore, the charges \eqref{Charge1} become
\begin{equation}
\mathcal{C}_{\xi}=-\int_{\Sigma_t}\text{d}^{d}x\left(\frac{\xi^t}{2}\frac{\Omega}{\sqrt{a}}\psi^2+\xi^i\psi\partial_i\phi\right).
\end{equation} 
These charges are expressed in Hamiltonian formalism. They are indeed conserved thanks to the equation of motion and together with the Poisson bracket they realize a representation of the Carrollian Killing algebra:
\begin{equation}
\{\mathcal{C}_{\xi},\mathcal{C}_{\eta}\}_{\text{Poisson}}=\int_{\Sigma_t}\text{d}^dx\left[\frac{\delta\mathcal{C}_{\eta}}{\delta\phi}\frac{\delta\mathcal{C}_{\xi}}{\delta\psi}-\frac{\delta\mathcal{C}_{\xi}}{\delta\phi}\frac{\delta\mathcal{C}_{\eta}}{\delta\psi}\right]=\mathcal{C}_{[\xi,\eta]}.
\end{equation}
This result confirms that the charges \eqref{Charge1} previously introduced are the correct ones. Finally, we notice that when $d=1$ the conformal state equation \eqref{traceless} is satisfied and the representation can be extended to conformal Killing vectors.

\section{Carrollian conservation laws in Ricci-flat gravity}\label{VI}

We will now turn our attention to Ricci-flat gravity. When the bulk metric is expressed in an appropriate gauge, usually given by imposing the radial coordinate be null, Einstein equations can reduce in some instances to equations defined on null infinity $\mathcal{I}^{+}$.\footnote{It will be the case for the three families of solutions we study in this section: the 3-dimensional asymptotically flat spacetimes, the weak field approximation of 4-dimensional asymptotically flat spacetimes in Bondi gauge and the Robinson Trautman solutions. The reduction of Einstein equations to equations on $\mathcal{I}^+$ would not be true, for example, for non-linearized 4-dimensional asymptotically flat gravity in Bondi gauge.} Its null nature makes it a natural host for a Carrollian geometry and the gravitational dynamics will be shown to match with Carrollian conservation laws. This section can be considered as a precursor of a full asymptotically flat holographic scheme. Indeed, the putative dual boundary theory would be Carrollian and live on $\mathcal{I}^{+}$. This theory would be coupled to a Carrollian geometry and satisfy Carrollian conservation laws that we map here to the gravitational dynamics. In gravity, the covariant phase-space formalism allows to compute surface charges, those will be shown to be given exactly or partially by the conserved charges defined in Sec. \ref{5.3}, depending whether the gravitational solution has radiation or not. To compute the charges explicitly, we use the code \cite{code}. 

\subsection{Asymptotically flat spacetimes in three dimensions}\label{6.1}

Three-dimensional asymptotically flat spacetimes are often studied in the Bondi gauge which, as we will shortly describe, imposes by definition the corresponding two-dimensional Carrollian manifold be flat. Here we want to show that we can source the geometric boundary fields, in order to create a general Carrollian structure \cite{Hartong2016}. 

Consider the following bulk metric
\begin{equation}
\text{d}s^2=g_{ab}\text{d}x^a\text{d}x^b=-2\text{u}\left(\text{d}r+r\left(\varphi_x\text{d}x+\theta\text{u}\right)\right)+r^2a_{xx}\text{d}x^2+8\pi G \text{u}\left(\mathcal{E}\text{u}-\pi_x\text{d}x\right).
\label{metric3d}
\end{equation}  
The bulk coordinates are $\{u,r,x\in\mathbf{S}^1\}$, $\text{u}=\Omega \text{d}u -b_x \text{d}x$, $a_{xx}$ is the one-dimensional boundary spatial metric, $\mathcal{E}$ and $\pi_x$ are the Carrollian momenta and $\theta$ and $\varphi_x$ correspond to \eqref{exp} and \eqref{acc} defined earlier:
\begin{equation}
\theta=\frac{1}{\Omega}\partial_u\ln\sqrt{a_{xx}} \quad\text{and}\quad\varphi_x=\frac{1}{\Omega}\left(\partial_x\Omega+\partial_ub_x\right).
\end{equation} 
All the fields appearing in the bulk metric depend only on $u$ and $x$. From this metric we can extract the corresponding Carrollian geometry on $\mathcal{I}^{+}=\{r\rightarrow\infty\}$. The following procedure is general but we will use the specific case of three-dimensional asymptotically flat spacetimes as an illustration. Consider the conformal extension of \eqref{metric3d}
\begin{equation}
\text{d}\tilde{s}^2=r^{-2}\text{d}s^2,
\end{equation}  
the factor $r^{-2}$ is present to regularize the metric on $\mathcal{I}^{+}$. We perform the change of variable $\omega=r^{-1}$ in the conformal metric, it becomes\footnote{The null asymptote is thus $\mathcal{I}^{+}=\{\omega\rightarrow 0\}$.}
\begin{equation}
\text{d}\tilde{s}^2=\tilde{g}_{ab}\text{d}x^a\text{d}x^b=-2\text{u}\left(-\text{d}\omega+\omega\left(\varphi_x\text{d}x+\theta\text{u}\right)\right)+a_{xx}\text{d}x^2+8\pi G\omega^2 \text{u}\left(\mathcal{E}\text{u}-\pi_x\text{d}x\right).
\end{equation}
We can deduce the Carrollian geometry on $\mathcal{I}^{+}$
\begin{equation}
\tilde{g}^{-1}\left(.,\text{d}\omega\right)_{\vert \mathcal{I}^{+}}=\frac{1}{\Omega}\partial_u,\quad \text{d}\tilde{s}^2_{\vert \mathcal{I}^{+}}=a_{xx}\text{d}x^2\quad\text{and}\quad \tilde{g}\left(.,\partial_{\omega}\right)_{\vert \mathcal{I}^{+}}=\Omega \text{d}u -b_i\text{d}x^i.
\end{equation}

We now move to the dynamics. In the following, we restrict our attention to the bulk line element \eqref{metric3d} with the additional geometrical constraint
\begin{equation}
\hat{\mathscr{D}}_xs^x=\hat{\nabla}_xs^x+2\varphi_xs^x=0,
\label{Constr}
\end{equation}
where $s_x=\frac{1}{\Omega}\partial_u\varphi_x-\theta\varphi_x-\hat{\partial}_x\theta$ is a Weyl-weight $1$ two-derivative object. The Carrollian momenta do not appear in this equation, it is just a constraint on the boundary geometrical background as it involves only the Carrollian metric fields. Using this ansatz, Einstein equations reduce to
\begin{eqnarray}
\left(\frac{1}{\Omega}\partial_u+2\theta \right)\mathcal{E} &=& 0, \label{Cons3dim1} \\
\left(\hat{\partial}_x+2\varphi_x\right)\mathcal{E}+\left(\frac{1}{\Omega}\partial_u+\theta\right)\pi_x &=& 0.
\label{Cons3dim2}
\end{eqnarray}
We interpret them as the Carrollian conservation equations \eqref{RC1}, \eqref{RC2}, \eqref{RC3} and \eqref{RC4} for $\Sigma^{xx}=\mathcal{B}^x=0$ and $\mathcal{E}=\mathcal{P}$ (conformal case). Furthermore $\Xi^{xx}$ is automatically zero due to its tracelessness. Therefore, the gravitational dynamics of this metric ansatz coincides with the Carrollian conservation equations that fall into the case described in Sec. \ref{5.3}.\footnote{With respect to Sec. \ref{5.3}, we trade here $t$ with $u$, to empathize that it is a retarded time.} 

{We would like at this point to obtain the surface charges. We thus compute the asymptotic Killing vectors of $\text{d}s^2$ whose leading orders in $r^ {-1}$ are
\begin{equation}
\hat{\xi^r}=-r\lambda(u,x)+\mathcal{O}(1),\quad \hat{\xi}^u=\xi^u(u,x)+\mathcal{O}(r^{-1})\quad \text{and}\quad\hat{\xi}^x=\xi^x(x)+\mathcal{O}(r^{-1}).
\end{equation} 
Here $\lambda=\hat{\nabla}_x\xi^x+\frac{X}{\Omega}\partial_u\ln\sqrt{a_{xx}}$ and $\xi=\xi^u\partial_u+\xi^x\partial_x$ is a conformal Killing vector (\emph{i.e.} satisfying \eqref{CCK1} and \eqref{CCK2}) of the corresponding Carrollian geometry $\{\Omega, a_{xx}, b_x\}$. We calculate the associated surface charge through covariant phase-space formalism and obtain that they are integrable and have exactly the same expression as the conserved charges defined in Sec. \ref{5.3} out of purely Carrollian considerations
\begin{equation}
Q_{\hat{\xi}}[\text{d}s^2]=\int_{\mathbf{S}^1}\text{d}x\sqrt{a_{xx}}\left(\left(\Omega\xi^u-2b_x\xi^x\right)\mathcal{E}-\xi^x\pi_x\right)=\mathcal{C}_{\xi}.
\end{equation}
There is no gravitational radiation in three dimensions, the charges are thus conserved. We will see that things are slightly different in four dimensions, where we have to consider the radiation at null infinity. 
 
If we restrict our attention to the case $\Omega=1$, $a_{xx}=1$ and $b_x=0$, we recover the usual Bondi gauge for asymptotically flat spacetimes and Carrollian conservation becomes 
\begin{eqnarray}
\partial_u\mathcal{E} &=& 0, \label{3dBondiCons1}\\
\partial_x\mathcal{E} &=& -\partial_u\pi_x.
\label{3dBondiCons2}
\end{eqnarray}
This set-up was extensively studied for instance in \cite{Barnich2012a}. Here, the solutions to the Carrollian Killing equations are exactly the bms$_3$ algebra vectors $\xi=\xi^u\partial_u+\xi^x\partial_x$ with $\xi^u=\partial_x\xi^xu+\alpha$, for any smooth functions $\xi^x(x)$ and $\alpha(x)$ on $\mathbf{S}^1$. Moreover the solutions to \eqref{3dBondiCons1} and \eqref{3dBondiCons2} are
\begin{equation}
\mathcal{E}(u,x)=\mathcal{E}_0(x)\quad\text{and}\quad \pi_x(u,x)=-\partial_x\mathcal{E}_0 u +\pi_0(x).
\end{equation} 
Hence, the charges become the usual ones
\begin{equation}
\mathcal{C}_{\xi}^{\text{Bondi}}=\int_{\mathbf{S}^1}\text{d}x\left(\alpha\mathcal{E}_0-\xi^x\pi_0\right),
\end{equation}
which are manifestly conserved. These were obtained in \cite{Barnich:2010eb, Barnich2012a}. \footnote{To compare, we have to identify $\phi=  x$, $\Xi(\phi)=-4\pi G \pi_{0}(x)$, $\Theta(\phi)=8\pi G \mathcal{E}_{0}(x)$, $Y(\phi)=\xi^x(x)$ and $ T(\phi)=\alpha(x)$.}

\subsection{Linearized gravity in four dimensions}\label{6.2}

We can perform the same kind of analysis in the case of asymptotically flat spacetimes in four dimensions, where asymptotic charges have been computed. We show that the boundary equations of motion, which are the linearized Einstein equations after gauge fixing, can be interpreted as a Carrollian conservation, and that the asymptotic charges are also charges associated with conformal Carrollian Killing vectors.

The bulk metric is $g_{ab}=\eta_{ab}+h_{ab}$ with
\begin{equation}
\begin{split}
&\eta=-\text{d}u^2-2\text{d}u\text{d}r+r^2\gamma_{ij}\text{d}x^i\text{d}x^j,\\
&h_{uu}=\frac{2}{r}m_B+\mathcal{O}\left(r^{-2}\right),\\
&h_{uj}=\frac{1}{2}\nabla^iC_{ij}+\frac{1}{r}N_j+\mathcal{O}\left(r^{-2}\right),\\
&h_{ij}=rC_{ij}+\mathcal{O}(1),\\
&h_{ra}=0.
\end{split}
\label{Bulk4D}
\end{equation}
where $a=\{r,\mu\}=\{r,u,x^i\}$, $i=1,2$. The perturbation $h_{ab}$ is traceless, so $\gamma^{ij}C_{ij}=0$, where $\gamma^{ij}$ is the metric of the two-sphere and $\nabla_i$ the associated covariant derivative. We recognize the mass aspect $m_B$, the angular momentum aspect $N_i$ and the gravitational wave aspect $C_{ij}$, all depending on $u$ and $x^i$. In this gauge, the linearized Einstein equations become:\footnote{Solving empty linearized Einstein equations order by order in $r^{-1}$ allows to express the various subleading coefficients in terms of $m_B$, $C_{ij}$ and $N_i$. The only residual equations are then the ones that we present here.}
\begin{eqnarray}
\partial_um_B &=& \frac{1}{4}\partial_ u \nabla^i\nabla^jC_{ij}, \label{Dynamics-Flat1}
\\
\partial_u N_i &=& \frac{2}{3}\partial_im_B-\frac{1}{6}\left[(\Delta-1)\nabla^jC_{ji}-\nabla_i\nabla^k\nabla^jC_{jk}\right].
\label{Dynamics-Flat2}
\end{eqnarray}
We first consider the case 
\begin{equation}
\nabla^i\nabla^jC_{ij}=0.
\end{equation} 
Then \eqref{Dynamics-Flat1} and \eqref{Dynamics-Flat2} admit a Carrollian interpretation and are recovered from \eqref{cons1} and \eqref{Cons2} with the following metric data
\begin{equation}
\Omega=1,\quad b_i=0,\quad a_{ij} = \gamma_{ij}, \label{AF4dData00}
\end{equation}
and Carrollian momenta
\begin{gather}
\Sigma^{ij}=\mathcal{B}^i = \Xi^i_i=0,\label{AF4dData01}\\
\mathcal{E}=4m_B,\quad\mathcal{A}^{ij} = -\frac{1}{2}\left(\frac{\mathcal{E}}{2}
a^{ij}-\Xi^{ij}\right),\quad\pi^i=-3N^i,\quad \Xi^i_j=\frac{1}{2}\left(\Delta-4\right) C^i_j,
\label{AF4dData}
\end{gather}
where $\mathcal{E}=-2\mathcal{A}^i_i$ and $\Xi^i_i=0$--we are in the conformal case. We obtain the following conservation equations:
\begin{eqnarray}
\partial_u\mathcal{E} &=& 0,\label{Cons-Flat1}\\
\partial_u\pi_i+\nabla_j\left(\frac{\mathcal{E}}{2}\gamma^j_i-\Xi^j_i\right) &=& 0.
\label{Cons-Flat2}
\end{eqnarray} 
This type of Carrollian conservation falls again into the class described in Sec. \ref{5.3}. 

The asymptotic Killing vectors $\hat{\xi}=\hat{\xi}^r\partial_r+\hat{\xi}^u\partial_u+\hat{\xi}^i\partial_i$ associated with the gauge \eqref{Bulk4D} have the following leading order in $r^{-1}$
\begin{equation}
\hat{\xi}^r=-\lambda(\mathbf{x})r+\mathcal{O}(1), \quad \hat{\xi}^u=\xi^u(t,\mathbf{x}) +\mathcal{O}(r^{-1})\quad\text{and}\quad\hat{\xi}^i=\xi^i(\mathbf{x})+\mathcal{O}(r^{-1}),
\end{equation}
where $\xi=\xi^u\partial_u+\xi^i\partial_i$ is a conformal Killing vector (\emph{i.e.} satisfying \eqref{CCK1} and \eqref{CCK2}) of the Carrollian geometry given by $\{\Omega=1, a_{ij}=\gamma_{ij}, b_i=0\}$ and $\lambda$ is the conformal factor. The solutions to the corresponding conformal Killing equations reproduce exactly the bms$_{4}$ algebra: $\xi^u=\frac{u}{2}\nabla_{i}\xi^i+\alpha(\mathbf{x})$, $\alpha$ being any function on $\mathbf{S}^2$, $\xi^i$ a conformal Killing of $\mathbf{S}^2$ and $\lambda=\frac{1}{2}\nabla_i\xi^i$.
We compute the corresponding surface charges. When $\nabla^i\nabla^jC_{ij}=0$ they take the form
\begin{equation}
Q_{\hat{\xi}}[g]=\int_{\mathbf{S}^2}\text{d}^2x\sqrt{\gamma}\left(\xi^u\mathcal{E}-\xi^i\pi_i\right)=\mathcal{C}_{\xi},
\label{Charge-Flat}
\end{equation}
with $\mathcal{E}$ and $\pi_i$ given by \eqref{AF4dData}. We recognize again the charges defined from purely Carrollian considerations in Sec. \ref{5.3}, associated with the data \eqref{AF4dData00}, \eqref{AF4dData01} and \eqref{AF4dData}. These charges are automatically conserved. Physically, this is due to the fact that part of the effect of gravitational radiation has suppressed by demanding $\nabla^i\nabla^jC_{ij}=0$. We will find shortly that relaxing this condition has an effect on the charge conservation.

Integrating \eqref{Cons-Flat1} and \eqref{Cons-Flat2} we obtain
\begin{equation}
\begin{split}
\mathcal{E}=\mathcal{E}_0(\mathbf{x}),\quad \pi_i=-\frac{1}{2}\partial_i\mathcal{E}_0u+\int \text{d}u^{\prime}\nabla_j\Xi^j_i+\pi_{0i}(\mathbf{x}).
\end{split}
\end{equation}
The charges become
\begin{equation}
\begin{split}
\mathcal{C}_{\xi}&=\int_{\mathbf{S}^2}\text{d}^2x\sqrt{\gamma}\left(\left(\frac{\nabla_i\xi^i}{2}u+\alpha\right)\mathcal{E}_0-\xi^i\left(-\frac{1}{2}\partial_i\mathcal{E}_0u+\int \text{d}u^{\prime}\nabla_j\Xi^j_i+\pi_{0i}\right)\right)\\
&=u\int_{\mathbf{S}^2}\text{d}^2x\sqrt{\gamma}\left(\frac{1}{2}\nabla_i(\xi^i\mathcal{E}_0)\right)+\int_{\mathbf{S}^2}\text{d}^2x\sqrt{\gamma}\left(\alpha\mathcal{E}_0-\xi^i\left(\int \text{d}u^{\prime}\nabla_j\Xi^j_i+\pi_{0i}\right)\right)\\
&=\int_{\mathbf{S}^2}\text{d}^2x\sqrt{\gamma}\left(\alpha\mathcal{E}_0-\xi^i\pi_{0i}\right)-\int \text{d}u^{\prime}\int_{\mathbf{S}^2}\text{d}^2x\sqrt{\gamma}\xi^i\nabla_j\Xi^j_i+\text{b.t.}\\
&=\int_{\mathbf{S}^2}\text{d}^2x\sqrt{\gamma}\left(\alpha\mathcal{E}_0-\xi^i\pi_{0i}\right)+\text{b.t.}.
\end{split}
\label{Charge4D}
\end{equation}
The  last step follows from the fact that $\xi^i$ is a conformal Killing vector on $\mathbf{S}^2$ and $\Xi^i_{j}$ is traceless. We observe that $\mathcal{C}_{\xi}$ is now manifestly conserved. 

When $\nabla^i\nabla^jC_{ij}\neq 0$, on the gravity side the radiation affects the surface charges and spoils their conservation. Therefore, these charges do not match those we defined earlier. This situation can be further investigated and recast in Carrollian language. To this end, we define $\sigma=\nabla^i\nabla^jC_{ij}$ and rewrite \eqref{Dynamics-Flat1} and \eqref{Dynamics-Flat2}
\begin{eqnarray}
\partial_u\mathcal{E} &=& 0,\label{Cons-Flat-Spoiled1}\\
\partial_u\pi_i+\nabla_j\left(\mathcal{P}\gamma^j_i-\Xi^j_i\right) &=& 0.
\label{Cons-Flat-Spoiled2}
\end{eqnarray}
Here, the metric fields are
\begin{equation}
\Omega=1,\quad b_i=0,\quad a_{ij} = \gamma_{ij},
\end{equation}
together with the Carrollian momenta
\begin{gather}
\Sigma^{ij}=\mathcal{B}^i =0,\\
\mathcal{E}=4m_B-\sigma,\quad\mathcal{P} =\frac{\mathcal{E}}{2}+\sigma ,\quad\pi^i=-3N^i,\quad \Xi^i_j=\frac{1}{2}\left(\Delta-4\right) C^i_j.
\end{gather}
Hence turning on $\sigma$ can be interpreted as spoiling the conformal state equation: $\mathcal{E}=-2\left(\mathcal{A}^i_i+\sigma\right)$. It appears as a sort of \emph{conformal anomaly} in the boundary theory. The surface charges become
\begin{equation}
Q_{\hat{\xi}}[g](u)=\int_{\mathbf{S}^2}\text{d}^2x\sqrt{\gamma}\left(\xi^u(\mathcal{E}+\sigma)-\xi^i\pi_i\right),
\label{Charge-Flat-Spoiled}
\end{equation}
and, as already stated, they are no longer conserved
\begin{equation}
\partial_u Q_{\hat{\xi}}[g]=\int_{\mathbf{S}^2}\text{d}^2x\sqrt{\gamma}\left(\delta_{\xi}+\lambda\right)\sigma,
\end{equation}
where $\delta_{\xi}$ is the usual Lie derivative and $\lambda=\frac{1}{2}\nabla_{i}\xi^i$ the conformal factor. These charges were obtained in \cite{Campoleoni2018}.\footnote{See the $n=2$ case of Sec. 3. Their charges coincide with \eqref{Charge-Flat-Spoiled} with $\alpha=T$, $\xi^i=v^i$, $\mathcal{E}_0=4\mathcal{M}$ and $\pi^i_0=-3\mathcal{N}^i$.} For non linear gravity see \cite{Barnich:2011mi}, where the charges are now non-integrable.

\subsection{Black hole solutions: Robinson--Trautman and Kerr--Taub--NUT}\label{6.3}

For asymptotically AdS solutions, Einstein equations lead to the conservation of an energy--momentum tensor on the timelike boundary with the cosmological constant playing the role of the velocity of light \cite{Ciambelli2018a}. Taking the flat limit in the bulk therefore corresponds to an ultra-relativistic limit on the boundary, and this is how Carrollian dynamics emerges. We illustrate this for the specific examples of Robinson--Trautman and Kerr--Taub--NUT, and analyze their charges.

\subsubsection*{Robinson--Trautman}

The Robinson--Trautman ansatz is
\begin{equation}
\text{d}s^2=\frac{2r^{2}}{P^2}\text{d}z\text{d}\bar{z}-2\text{d}u\text{d}r-\left(\Delta \ln P-2r\partial_u\ln P-\frac{2m}{r}\right)\text{d}u^2, \label{RT}
\end{equation}
where $m$ and $P$ depend on the boundary coordinates $\{u,z,\bar{z}\}$. This metric is Ricci-flat provided the Robinson--Trautman equations are satisfied:
\begin{eqnarray}
\Delta\Delta\ln P+12 M\partial_u\ln P-4\partial_u M &=& 0,\\
\partial_ zM &=& 0,\\
\partial_{\bar{z}}M &=& 0,
\label{RTeq}
\end{eqnarray}
where we have defined $\Delta=\nabla^i\nabla_i$, for $i=\{z,\bar{z}\}$, and $\nabla_{i}$ is the Levi Civita covariant derivative of the spatial metric $a=\frac{2}{P^2}\text{d}z\text{d}\bar{z}$. These equations can be interpreted as Carrollian conservation laws \eqref{RC1}, \eqref{RC2}, \eqref{RC3} and \eqref{RC4} with the metric data $\Omega=1$, $b_i=0$ and $a=\frac{2}{P(u,z,\bar{z})^2}\text{d}z\text{d}\bar{z}$ and the Carrollian momenta
\begin{gather}
\Xi^{ij}=\pi^i=\Sigma^i_i=0,\\
\mathcal{E}=4 M,\quad \mathcal{B}^i=\nabla^iK,\quad\mathcal{A}^{ij}=-Ma^{ij},\quad \Sigma^{ij}=\nabla^i\nabla^j\theta-\frac{1}{2}a^{ij}\nabla^k\nabla_k\theta.
\end{gather} 
Here we have introduced the Gaussian curvature $K=\Delta\ln P$. Weyl covariance is ensured by the conformal state equation $\mathcal{E}=-2\mathcal{A}^i_i$, together with $\Sigma^i_i=0$. With this set of data, the conservation equations are
\begin{eqnarray}
\left(\partial_u+\frac{3\theta}{2}\right)\mathcal{E}-\nabla_i\mathcal{B}^i &=& 0, \label{RTRC1}\\
\partial_j\mathcal{E} &=& 0, \label{RTRC2}\\
\left(\partial_u+\theta\right)\mathcal{B}_j+\nabla_i\Sigma^i_j &=& 0, \label{RTRC3}\\
\Sigma^{ij}\xi_{ij}+\frac{\theta}{d}\Sigma^i_i &=& 0. 
\label{RTRC4}
\end{eqnarray}
Equations \eqref{RC3} and \eqref{RC4} do not appear in the Robinson--Trautman equations because they are geometrical constraints on the spatial metric, which are automatically satisfied when imposing $a=\frac{2}{P^2}\text{d}z\text{d}\bar{z}$.

We want to interpret the charges we have introduced in Secs. \ref{5.2} and \ref{5.3} for the Robinson--Trautman spacetime. To this end, we introduce a conformal Carrollian Killing vector $\xi$, with \eqref{CCK1} and \eqref{CCK2} here given by
\begin{eqnarray}
\partial_u\xi^u &=& \lambda,\label{CCKRT2}\\
\nabla_i\xi_j+\nabla_j\xi_i+\xi^u\partial_ua_{ij} &=& 2\lambda a_{ij}. \label{CCKRT1}
\end{eqnarray}
The solution is the following vector\footnote{The metric \eqref{RT} is not in the Bondi gauge unless $P$ is time independent. Therefore, the conformal Killing vector $\xi$ does not satisfy the usual bms$_4$ algebra, but a generalized version of it.}
\begin{equation}
\xi=\left(\sqrt{a}\right)^{\frac{1}{2}}\left(\alpha(\mathbf{x})+\frac{1}{2}\int \text{d}u \left(\sqrt{a}\right)^{-\frac{1}{2}}\nabla_{i}\xi^i\right)\partial_u+\xi^i(\mathbf{x})\partial_i,
\end{equation}
where $\xi^i$ is a spatial conformal Killing vector, \emph{i.e.} it satisfies
\begin{equation}
\nabla_{i}\xi_j+\nabla_{j}\xi_{i}=\nabla_{k}\xi^ka_{ij}.
\end{equation}
The associated charges \eqref{charge2} become
\begin{equation}
\mathcal{Q}_{\xi}=\int_{\mathbf{S}^2} \text{d}^2z\sqrt{a}\xi_{j}\mathcal{B}^j=\int_{\mathbf{S}^2} \text{d}^2z P^{-2}\left(\xi^{z}\partial_zK+\xi^{\bar{z}}\partial_{\bar{z}}K\right).
\end{equation}
They are conserved by construction. 

Even though the second family of charges \eqref{Charge1} were defined only for ${\cal B}^i=0$, we can nevertheless study what their expression is for the solution at hand. We find
\begin{equation}
\mathcal{C}_{\xi}=\int_{\mathbf{S}^2} \text{d}^2z\sqrt{a}\xi^u\mathcal{E}=\int_{\mathbf{S}^2} \text{d}^2z P^{-3}\left(\alpha(z,\bar{z})+\frac{1}{2}\int \text{d}u P\nabla_i\xi^i\right)4M.
\end{equation}
As expected, they are not generically conserved, and using \eqref{RTRC1} we find
\begin{equation}
\partial_u\mathcal{C}_{\xi}=-\int_{\mathbf{S}^2} \text{d}^2z\sqrt{a}\partial_i\xi^u\mathcal{B}^i.
\end{equation}
Their conservation holds in two instances. The first, expected by construction, is when ${\cal B}_i=\partial_iK=0$, and corresponds to a uniform curvature of the boundary sphere at all times. The second, which is a new condition, occurs when the conformal Killing vectors satisfy also $\partial_i\xi^u=0$. This can be written as
\begin{equation}
\delta_{\xi}b_i=0,
\end{equation}
when considering the Robinson--Trautman Carrollian geometry $\Omega=1$, $b_i=0$ and $a=\frac{2}{P^2}\text{d}z\text{d}\bar z$.\footnote{Actually, it is possible to show that, even when $\mathcal{B}^i\neq 0$, the charges \eqref{Charge1} are generically conserved if the vectors $\xi$ satisfy $\delta_{\xi}a_{ij}=0$, $\delta_{\xi}\Omega=0$ \emph{and} $\delta_{\xi}b_i=0$.}

\subsubsection*{Kerr--Taub--NUT family}

The interesting feature of the Kerr--Taub--NUT family is that, although stationary, it has a non-trivial metric field $b_i$. Its line element, in $\{t,r,\theta,\phi\}$ coordinates, is given by
\begin{equation}
\text{d}s^2=-\dfrac{\Delta_r}{\rho^2}\left(\text{d}t-\text{b}\right)^2+\dfrac{\rho^2}{\Delta_r}\text{d}r^2+\rho^2\left(\text{d}\theta^2+\sin^2\theta \text{d}\phi^2\right)+\dfrac{\sin^2 \theta}{\rho^2}\left(\alpha \text{d}t-\left(r^2+(n-\alpha)^2\right)\text{d}\phi\right)^2,
\end{equation}
where
\begin{eqnarray}
\Delta_r &=& -2 M r+r^2+\alpha^2-n^2, \\
\rho^2 &=& r^2+(n-\alpha \cos\theta)^2, \\
\text{b} &=& \left(2n (\cos\theta-1)+\alpha \sin^2\theta\right) \text{d}\phi. \label{bkerr}
\end{eqnarray}
In this solution, $M$ is interpreted as the black hole mass, $\alpha$ its angular parameter and $n$ its NUT charge. The Carrollian geometrical data are $\Omega=1$, $b_i$ as in \eqref{bkerr} and $a=\text{d}\theta^2+\sin^2\theta \text{d}\phi^2$. The bulk Einstein equations are satisfied for a constant mass. We can interpret this result as given by the following Carrollian data
\begin{equation}
\Xi^{ij}=\pi^i=\Sigma^{ij}={\cal B}^i=0 \quad {\cal E}=M \quad {\cal A}^{ij}=-\dfrac{M}{4}a^{ij},
\end{equation}
such that Carrollian conservation equations give straightforwardly $M$ constant. From the hydrodynamical viewpoint, these data describe a perfect fluid. 

The conformal Carrollian Killing equations can be solved with the result
\begin{equation}
\xi=\left(T(\mathbf{x})+\frac{1}{2} t \nabla_{i}\xi^i\right)\partial_t+\xi^i(\mathbf{x})\partial_i.
\end{equation}
where $T$ is any smooth function on $\mathbf{S}^2$ and $\xi^i$ a Killing vector of the sphere. This is precisely the bms$_4$ generator. The charges \eqref{charge2} are identically zero in this case. Conversely, the charges \eqref{Charge1} are non-trivial
\begin{equation}
{\cal C}_{\xi}=M\int_{\mathbf{S}^2}\text{d}\theta \text{d}\phi \sin\theta\left(T-\dfrac{3}{2}\xi^ib_i\right).
\end{equation}
They explicitly depend on the Kerr--Taub--NUT parameters thanks to the presence of the metric field $b_i$, and they are manifestly conserved.

\newpage

\section{Conclusions}

We are now ready to summarize our achievements. 

In the framework of Carrollian dynamics we have defined Carrollian momenta as the variation of the action with respect to the Carrollian metric fields $\Omega, b_i, a_{ij}$.  These momenta obey conservation laws ensuing the invariance of the action under Carrollian diffeomorphisms. We have carefully stressed that this set of Carrollian momenta plays the role the energy--momentum tensor has in relativistic theories, since such an object cannot be defined in general Carrollian dynamics. In the very particular instance of flat Carrollian geometry, due to the existence of global symmetries, the on-shell Carrollian momenta are indistinguishable from the N\oe ther conserved currents. In this case they can be packaged in a non-symmetric spacetime energy--momentum tensor. 

We have proven that the general conservation equations of the set of Carrollian momenta are recovered as the ultra-relativistic limit of the relativistic energy--momentum tensor conservation equations. This is expected and shows in passing that the Carrollian limit of the energy--momentum tensor outside its conservation equations is non sensible. 

As usual in theories with local symmetries, volume conserved charges cannot be defined from plain conserved momenta. Killing fields are needed, in order to construct conserved currents and extract conserved charges, which encode the physical information stored in the fields at hand. We performed all these steps in a general Carrollian geometry, starting with the definition of the  Killing vectors and proceeding with currents (projections of the Carrollian momenta) and charges. 

All these concepts and techniques have been finally illustrated in concrete examples inspired from flat holography. Indeed, the null infinity of an asymptotically flat spacetime is a natural host for Carrollian geometry, and Carrollian conservation equations on ${\cal I}^+$ emerge as part of the bulk Einstein dynamics. More specifically, we have shown that in three bulk dimensions the Carrollian charges match the surface charges obtained from standard bulk methods. However, in four-dimensional linearized gravity, the presence of gravitational radiation spoils the conservation of surface charges. At the level of the Carrollian conservation equations, this is interpreted as a conformal anomaly, the radiation sourcing the anomalous factor. The subsequent analysis of the Robinson--Trautman  and Kerr--Taub--NUT exact solutions nicely confirms these expectations and the interplay among the bulk and the boundary dynamics.

Our analysis triggers many questions. Among others, the two examples of exact Ricci-flat spacetimes 
treated here suggest to further investigate the Carrollian interpretation of four-dimensional gravity in full generality, \emph{i.e.} without assuming linearity. More generally, this work may help in paving the road toward the Carrollian understanding of flat holography, already discussed in several instances in the literature.

\section*{Acknowledgments}

We would like to thank Glenn Barnich, Quentin Bonnefoy, Guillaume Bossard, Andrea Campoleoni, Thibault Damour, Laura Donnay, Francesco Galvagno, Monica Guica, Rob Leigh, Rodrigo Olea, Tassos Petkou, Ana-Maria Raclariu, Kostas Siampos, Piotr Tourkine and Sasha Zhiboedov for useful discussions. We are particularly grateful to Marios Petropoulos for carefully reading the final manuscript. Luca Ciambelli would like to acknowledge the University of Torino and Milano Bicocca where part of this work was developed. This work was partly funded by the ANR-16-CE31-0004 contract Black-dS-String.

\appendix

\section{Carrollian Charges algebra}\label{App}
We have defined two types of conserved charges in \ref{5.2} and \ref{5.3},  $\mathcal{Q}_{\xi}$ and $\mathcal{C}_{\xi}$.  The first one is conserved for any type of Carrollian conservation laws given by \eqref{RC1}, \eqref{RC2}, \eqref{RC3} and \eqref{RC4}, while the second is conserved only when the Carrollian momenta $\mathcal{B}^i$ vanishes. We recall their expression:
\begin{equation}
\mathcal{Q}_{\xi}=\int_{\Sigma_t}\text{d}^dx\sqrt{a}\xi_i\left(\mathcal{B}^i+b_ j\Sigma^{ji}\right)\quad\text{and}\quad {\cal C}_{\xi}=\int_{\Sigma_t}\text{d}^dx\sqrt{a}\left(X\mathcal{E}-\xi^i\pi_i+2b_i\xi^j\mathcal{A}^i_j\right).
\end{equation}
In this appendix we show that both of them are also representations of the (conformal) Carrollian Killing algebra.

Consider two Carrollian Killing vectors $\xi$ and $\eta$. It is possible to decompose them in a coordinate basis,
\begin{equation}
\xi=\xi^t(t,\mathbf{x})\partial_t+\xi^i(\mathbf{x})\partial_i\quad\text{and}\quad\eta=\eta^t(t,\mathbf{x})\partial_t+\eta^i(\mathbf{x})\partial_i,
\end{equation}
or in a Carroll-covariant one,
\begin{equation}
\xi=\frac{X}{\Omega}\partial_t+\xi^i\hat{\partial}_i\quad\text{and}\quad\eta=\frac{Y}{\Omega}\partial_t+\eta^i\hat{\partial}_i,
\end{equation}
where $X=\Omega\xi^t-b_i\xi^i$, $Y=\Omega\eta^t-b_i\eta^i$ and $\hat{\partial}_i$ is the Carroll-covariant spatial derivative defined in \ref{2.2}. The commutator of $\xi$ and $\eta$ is given by 
\begin{equation}
\begin{split}
\lambda\equiv[\xi,\eta]&=\left(\xi^t\partial_t\eta^t-\eta^t\partial_t\xi^t+\xi^k\partial_k\eta^t-\eta^k\partial_k\xi^t\right)\partial_t+\left(\xi^k\partial_k\eta^i-\eta^k\partial_k\xi^i\right)\partial_i=\frac{L}{\Omega}\partial_t+\lambda^i\hat{\partial}_i.
\end{split}
\end{equation}
For $\xi$ and $\eta$ Carrollian Killing vectors, we define the two following quantities
\begin{equation}
\begin{split}
&\{\mathcal{Q}_{\xi},\mathcal{Q}_{\eta}\}\equiv\int_{\Sigma_t} \text{d}^dx\delta_{\eta}\left[\sqrt{a}\xi_i\left(\mathcal{B}^i+b_ j\Sigma^{ji}\right)\right],\\
&\{\mathcal{C}_{\xi},\mathcal{C}_{\eta}\}\equiv\int_{\Sigma_t} \text{d}^dx\delta_{\eta}\left[\sqrt{a}\left(X\mathcal{E}-\xi^i\pi_i+2b_i\xi^j\mathcal{A}^i_j\right)\right],
\end{split}
\end{equation}
where $\delta_{\eta}$ is the Lie derivative w.r.t. $\eta$ acting on the metric fields and the momenta, but not on $\xi^t$ and $\xi^i$. We want to show that, up to boundary terms,
\begin{equation}
\{\mathcal{Q}_{\xi},\mathcal{Q}_{\eta}\}=\mathcal{Q}_{[\xi,\eta]}\quad\text{and}\quad\{\mathcal{C}_{\xi},\mathcal{C}_{\eta}\}=\mathcal{C}_{[\xi,\eta]},
\end{equation}
the first result being true for any type of Carrollian conservation laws while the second one holds only when $\mathcal{B}^i=0$.

We start with the first one, we have
\begin{equation}
\begin{split}
\{\mathcal{Q}_{\xi},\mathcal{Q}_{\eta}\}=&\int_{\Sigma_t} \text{d}^dx\bigg[\delta_{\eta}\sqrt{a}\xi_i\left(\mathcal{B}^i+b_ j\Sigma^{ji}\right)+\sqrt{a}(\delta_{\eta}a_{ik})\xi^k\left(\mathcal{B}^i+b_ j\Sigma^{ji}\right)\\
&+\sqrt{a}\xi_i\left(\delta_{\eta}\mathcal{B}^i+\delta_{\eta}b_ j\Sigma^{ji}+b_ j\delta_{\eta}\Sigma^{ji}\right)\bigg].
\end{split}
\end{equation}
We compute the infinitesimal variations of the geometric fields and the Carrollian momenta:
\begin{eqnarray}
\delta_{\eta}a_{ik} &=& \eta^t\partial_ta_{ik}+\eta^j\partial_ja_{ik}+\partial_i\eta^ja_{kj}+\partial_k\eta^ja_{ij}=0,\\
\delta_{\eta}\sqrt{a} &=& \eta^i\partial_i\sqrt{a}+\eta^t\partial_t\sqrt{a}+\partial_i\eta^i\sqrt{a}=0,\\
\delta_{\eta}b_i &=& \eta^t\partial_tb_i+\eta^j\partial_jb_i-\Omega\partial_i\eta^t+b_j\partial_i\eta^j,\\
\delta_{\eta}\mathcal{B}^i &=& \eta^t\partial_t\mathcal{B}^i+\eta^j\partial_j\mathcal{B}^i-\mathcal{B}^j\partial_j\eta^i,\\
\delta_{\eta}\Sigma^{ij} &=& \eta^t\partial_t\Sigma^{ij}+\eta^k\partial_k\Sigma^{ij}-\Sigma^{kj}\partial_k\eta^i-\Sigma^{ik}\partial_k\eta^j.
\end{eqnarray}
The variation of $a_{ik}$ and $\sqrt{a}$ vanish because $\eta$ is a Carrollian Killing vector. Then we eliminate every temporal derivative of the Carrollian momenta using the conservation laws \eqref{RC3} and \eqref{RC4}. Finally performing integration by parts and using properties of the Carrollian Killing vectors \eqref{CK1} and \eqref{CK2}, we suppress every spatial derivative of the Carrollian momenta to obtain:
\begin{equation}
\{\mathcal{Q}_{\xi},\mathcal{Q}_{\eta}\}=\int_{\Sigma_t}\text{d}^dx\sqrt{a}\lambda_i\left(\mathcal{B}^i+b_ j\Sigma^{ji}\right)+\text{b.t.}=\mathcal{Q}_{\lambda}+\text{b.t.}.
\end{equation}
This proves that the charges $\mathcal{Q}_{\xi}$ form a representation of the Carrollian Killing algebra.

We now prove the second relation. We have
\begin{equation}
\begin{split}
\{\mathcal{C}_{\xi},\mathcal{C}_{\eta}\}=&\int_{\Sigma_t}\text{d}^dx\bigg[\delta_{\eta}\sqrt{a}\left((\Omega\xi^t-b_i\xi^i)\mathcal{E}-\xi^i\pi_i+2b_i\xi^j\mathcal{A}^i_j\right)\\
&+\sqrt{a}\left((\delta_{\eta}\Omega\xi^t-\delta_{\eta}b_i\xi^i)\mathcal{E}+(\Omega\xi^t-b_i\xi^i)\delta_{\eta}\mathcal{E}-\xi^i\delta_{\eta}\pi_i+2\delta_{\eta}b_i\xi^j\mathcal{A}^i_j+2b_i\xi^j\delta_{\eta}\mathcal{A}^i_j\right)\bigg].
\end{split}
\end{equation}
We compute the infinitesimal variations of the geometric fields and the Carrollian momenta:
\begin{eqnarray}
\delta_{\eta}\Omega &=& \eta^t\partial_t\Omega+\eta^i\partial_i\Omega+\Omega\partial_t\eta^t=0,\\
\delta_{\eta}\sqrt{a} &=& \eta^i\partial_i\sqrt{a}+\eta^t\partial_t\sqrt{a}+\partial_i\eta^i\sqrt{a}=0,\\
\delta_{\eta}b_i &=& \eta^t\partial_tb_i+\eta^j\partial_jb_i-\Omega\partial_i\eta^t+b_j\partial_i\eta^j,\\
\delta_{\eta}\mathcal{E} &=& \eta^i\partial_i\mathcal{E}+\eta^t\partial_t\mathcal{E},\\
\delta_{\eta}\pi_i &=& \eta^t\partial_t\pi_i+\eta^j\partial_j\pi_i+\pi_j\partial_i\eta^j,\\
\delta_{\eta}\mathcal{A}^i_j &=& \eta^t\partial_t\mathcal{A}^i_j+\eta^k\partial_k\mathcal{A}^i_j-\mathcal{A}^k_j\partial_k\eta^i+\mathcal{A}^i_k\partial_j\eta^k.
\end{eqnarray}
The variations of $\Omega$ and $\sqrt{a}$ are vanishing because $\eta$ is a Carrollian Killing vector. Then we eliminate every temporal derivative of the Carrollian momenta using the conservation laws \eqref{RC11} and \eqref{RC22}. Finally performing integration by parts and using properties of the Carrollian Killings, \eqref{CK1} and \eqref{CK2}, we suppress every spatial derivative of the Carrollian momenta to obtain:
\begin{equation}
\begin{aligned}
\{\mathcal{C}_{\xi},\mathcal{C}_{\eta}\}=&\int_{\Sigma_t} \text{d}^dx\sqrt{a}\bigg[\left(\Omega(\xi^t\partial_t\eta^t-\eta^t\partial_t\xi^t+\xi^k\partial_k\eta^t-\eta^k\partial_k\xi^t)-b_i(\xi^k\partial_k\eta^i-\eta^k\partial_k\xi^i)\right)\mathcal{E}\\
&-(\xi^k\partial_k\eta^i-\eta^k\partial_k\xi^i)\pi_i+2b_i(\xi^k\partial_k\eta^j-\eta^k\partial_k\xi^j)\mathcal{A}^i_j\bigg]+\text{b.t.},
\end{aligned}
\end{equation}
which corresponds to 
\begin{equation}
\{\mathcal{C}_{\xi},\mathcal{C}_{\eta}\}=\int_{\Sigma_t}\text{d}^dx\sqrt{a}\left(L\mathcal{E}-\xi^i\pi_i+2b_i\lambda^j\mathcal{A}^i_j\right)+\text{b.t.}=\mathcal{C}_{\lambda}+\text{b.t.}.
\end{equation}
Therefore, up to boundary terms, the charges $\mathcal{C}_{\xi}$ form a representation of the Carrollian Killing algebra.

We can extend the previous results to the conformal Carrollian Killing algebra when imposing $\Sigma^i_i=0$ and the conformal state equation $\mathcal{E}=-2\mathcal{A}^i_i$.

\bibliography{Refs}
\bibliographystyle{utphys}
\end{document}